%% file: BAM764_review.tex
\documentclass[prd,aps,twocolumn,showpacs,amsmath,amssymb]{revtex4-2}
\usepackage{subfigure}
\usepackage{graphicx}
\usepackage{graphicx}
\usepackage{array}
\usepackage{xcolor}

\usepackage{lineno}
\usepackage{bm}
\usepackage{multirow}
\usepackage{makecell}
\usepackage[colorlinks,linkcolor=blue,anchorcolor=blue,citecolor=blue,urlcolor=blue]{hyperref}

\begin{document}
\normalsize
\parskip=5pt plus 1pt minus 1pt


\title{Measurements of the $CP$-even fractions of $D^0\to\pi^{+}\pi^{-}\pi^{0}$ and $D^0\to K^{+}K^{-}\pi^{0}$ at BESIII}

\input{./authorlist_2024-04-30.tex}

\begin{abstract}

The $CP$-even fractions ($F_{+}$) of the decays $D^0\to\pi^{+}\pi^{-}\pi^{0}$ and $D^0\to K^{+}K^{-}\pi^{0}$ are measured with a quantum-correlated $\psi(3770)\to D\bar{D}$ data sample collected by the BESIII experiment corresponding to an integrated luminosity of 7.93 $\mathrm{fb}^{-1}$. The results are  $F_{+}^{\pi^{+}\pi^{-}\pi^{0}}=0.9406\pm0.0036\pm0.0021$ and $F_{+}^{K^{+}K^{-}\pi^{0}}=0.631\pm0.014\pm0.011$, where the first uncertainties are statistical and the second systematic. 
These measurements are consistent with the previous determinations, and the uncertainties for $F_{+}^{\pi^{+}\pi^{-}\pi^{0}}$ and $F_{+}^{K^{+}K^{-}\pi^{0}}$ are reduced by factors of 3.9 and 2.6, respectively. The reported results provide important inputs for the precise measurement of the angle $\gamma$ of the Cabibbo-Kobayashi-Maskawa matrix and indirect $CP$ violation in charm mixing.

\end{abstract}

\maketitle

\section{INTRODUCTION}

The Cabibbo-Kobayashi-Maskawa (CKM) matrix, which is one of the most important components of the standard model (SM), characterizes the coupling strength between up and down quarks through weak interactions and its complex phase is the only source of $CP$ violation in the SM~\cite{BSM}. The unitarity triangle is a graphical representation of the CKM matrix in the complex plane, and include the three angles denoted as $\alpha$, $\beta$ and $\gamma$~\cite{PDG2023}. 
Comparison of determinations of $\gamma$ from direct measurements and that from the global CKM fit provides an important test of the CKM unitarity and is sensitive to new physics beyond the SM.

Experimentally, $\gamma$ can be extracted through the golden decay mode of $B^{\pm}\to DK^{\pm}$~\cite{Gronau1991}, where the $D$ is a superposition of $D^{0}$ and $\bar{D}^{0}$.  The $D$ meson can be reconstructed in decays of mixed $CP$ content~\cite{CLEO2015a}, quantified by the $CP$-even fraction~($F_{+}$).
In practice, the quantum correlated $\psi(3770)\to D\bar{D}$ data sample provides a unique platform to measure $CP$-even fraction and other parameters associated with the strong dynamics of neutral $D$ decays. 
Previous measurements of $CP$-even fractions have been made with data collected by  the CLEO-c experiment, including $D^0\to \pi^{+}\pi^{-}\pi^{0}$ and $D^0\to K^{+}K^{-}\pi^{0}$~\cite{CLEO2015a,CLEO2015b}, and the BESIII experiment, including $D^0\to \pi^+\pi^-\pi^{+}\pi^{-}$, $D^0\to K^{+}K^{-}\pi^{+}\pi^{-}$ and $D^0\to K_{S}^{0}\pi^{+}\pi^{-}\pi^{0}$~\cite{BESIII:Fp4pi,BESIII:Fp2K2pi,BESIII:FpKs3pi}. 

The $CP$-even fractions of $D^0\to\pi^{+}\pi^{-}\pi^{0}$ ($F_{+}^{ \pi^+\pi^{-}\pi^{0}}$) and $D^0\to K^{+}K^{-}\pi^{0}$ ($F_{+}^{ K^+K^{-}\pi^{0}}$) are two  important input decay parameters in measuring the $\gamma$ angle through the decay $B^{\pm}\to DK^{\pm}$~\cite{Aaij2021} and the indirect $CP$ violation in charm mixing~\cite{LHCb2016a}. With the large $B$ meson samples expected from the LHCb and Belle-II experiments in the coming years~\cite{LHCb2019a,BelleII2019a}, the uncertainties of these $CP$-even fractions will become a dominant source of uncertainty and the bottleneck in the future precise measurement of $\gamma$ with these decay channels. Therefore, improving the precision of these $CP$-even fractions is highly desirable, exploiting the large  $\psi(3770)$ data sample collected with the BESIII experiment~\cite{BESIII2020a}. This paper presents the measurements of the $CP$-even fractions $F_{+}^{ \pi^+\pi^{-}\pi^{0}}$ and $F_{+}^{ K^+K^{-}\pi^{0}}$ using a quantum-correlated $\psi(3770)\to D\bar{D}$ data sample collected in the BESIII experiment, corresponding to an integrated luminosity of 7.93 $\mathrm{fb}^{-1}$~\cite{BESIII:NDD1,BESIII:NDD2}.

\section{MEASUREMENT METHOD}

The wave function for the ${e^{+}e^{-}\to \psi(3770)\to D^{0}\bar{D}^{0}}$ process
is asymmetric due to the odd charge conjugation of the  $\psi(3770)$. The quantum coherence of the pair of charm mesons    provides a unique opportunity to measure the $CP$-even fractions of neutral $D$ decays using a double-tag (DT) method~\cite{MarkIII:DT}. In this method, the $D$ meson is reconstructed with the signal-decay modes of interest, denoted as $g$ hereafter, while the $\bar{D}$ meson is reconstructed by any of several tag modes with different $CP$ states, denoted as $f$.  The signal modes of interest can themselves be used as tags.  The tag modes are also reconstructed in  single-tag (ST) samples, where no requirement is placed on the decay of the other $D$ meson in the event.  It is not possible, however, to reconstruct ST samples for modes involving a $K^0_L$ meson.
The tag modes used in this analysis are categorized into different $CP$ types with full or binned phase space as summarized in Table~\ref{tab:tagmode} and the inclusion of charge conjugate modes is implicit throughout this paper.

\begin{table}[htbp]
	\centering
	\caption{The tag modes used, categorised by $CP$ content.} 
	\label{tab:tagmode}	
	\begin{tabular}[b]{ p{2cm} p{1.5cm}  p{1.5cm} }
		\hline
		\hline
		\multicolumn{1}{c}{\small{Type}} &  \multicolumn{2}{c}{\small{Tag mode}} \\
		\hline
		\small{$CP$-even}                          &  \multicolumn{2}{c}{$K^{+}K^{-}$, $\pi^{+}\pi^{-}$, $K^{0}_{S}\pi^{0}\pi^{0}$, $K^{0}_{L}\omega_{\pi^{+}\pi^{-}\pi^{0}}$, $K^{0}_{L}\pi^{0}$} \\
		\hline
		\multirow{2}*{\small{$CP$-odd}} &     \multicolumn{2}{c}{$K^{0}_{S}\pi^{0}$, $K^{0}_{S}\eta_{\gamma\gamma}$, $K^{0}_{S}\eta_{\pi^{+}\pi^{-}\pi^{0}}$, $K^{0}_{S}\eta'_{\gamma\rho^{0}}$,} \\
                                &     \multicolumn{2}{c}{$K^{0}_{S}\eta'_{\pi^{+}\pi^{-}\eta}$, $K^{0}_{S}\omega_{\pi^{+}\pi^{-}\pi^{0}}$, $K^{0}_{L}\pi^{0}\pi^{0}$} \\
		\hline
		\multicolumn{1}{l}{\small{$CP$-mixed }}                  &   \multicolumn{2}{c}{\multirow{2}{*}{$K^{+}K^{-}\pi^{0}$, $\pi^{+}\pi^{-}\pi^{0}$, $\pi^{+}\pi^{-}\pi^{+}\pi^{-}$}} \\
		\multicolumn{2}{l}{\small{full phase space}} &  \\
		\hline
		\multicolumn{1}{l}{\small{$CP$-mixed }}    &  \multicolumn{2}{c}{\multirow{2}{*}{$K^{0}_{S,L}\pi^{+}\pi^{-}$}} \\
		\multicolumn{2}{l}{\small{binned phase space}}  &  \\
			
		\hline	
		\hline
	\end{tabular}
\end{table}

With the above approach, for the $CP$ eigen (including $CP$-even and $CP$-odd) ST mode $f$, the expected ST yield is given by
\begin{equation} 
	\label{eq:NST_PureCP}
	S(f)=2N_{D\bar{D}}\mathcal{B}(f)\epsilon_{\mathrm{ST}}^{f}[1-\eta^{f}_{CP}y],
\end{equation}
\noindent
where $N_{D\bar{D}}$ is the total number of $D\bar{D}$ pairs in the used data sample, $\mathcal{B}(f)$ and $\epsilon_{\mathrm{ST}}^{f}$ are the corresponding branching fraction and reconstruction efficiency of the ST mode $f$, $\eta^{f}_{CP}$ is its $CP$ eigenvalue ($\pm 1$), and $y$ is the charm-mixing parameter $y=(0.645^{+0.024}_{-0.023})\%$~\cite{HFLAV:y}.
Meanwhile the excepted DT yield can be written as~\cite{CLEO2015a}
\begin{equation} 
	\label{eq:NDT_PureCP}
	M(g,f)=2N_{D\bar{D}}\mathcal{B}(g)\mathcal{B}(f)\epsilon_{\mathrm{DT}}^{g,f}[1-\eta^{f}_{CP}(2F^{g}_{+}-1)],
\end{equation}
\noindent
where $\mathcal{B}(g)$ is the branching fraction of the signal mode, $\epsilon_{\mathrm{DT}}^{g,f}$ is the DT reconstruction efficiency and $F^{g}_{+}$ is its $CP$-even fraction.  
Terms of $\mathcal{O}(y^{2})$ or higher are neglected. 

The $CP$-even fraction $F^{g}_{+}$ is defined as
\begin{equation} 
	\label{eq:Fp_PureCP}
	F^{g}_{+}=\frac{N^{+}}{N^{+}+N^{-}},
\end{equation}
\noindent
where $N^{+}$ ($N^{-}$) is the decay rate of $CP$-even (odd) $D$ mesons for the signal mode, which can be obtained from the $CP$-odd (even) ST $\bar{D}$ sample. That is 
\begin{equation}
N^{\pm} =  \frac{M(g,f)[1\pm y]}  {S(f)\epsilon_{\mathrm{DT}}^{g,f}/\epsilon_{\mathrm{ST}}^{f}}.
\label{eq:eq4}
\end{equation}

In this analysis, for the tag modes not involving a  $K_{L}^0$, both ST and DT candidates are reconstructed fully (referred to fully reconstructed events thereafter). Therefore $N^{\pm}$ can be calculated using the measured ST and DT yields, along with the corresponding efficiencies estimated from Monte Carlo (MC) samples.
However, for the tag modes involving a $K_{L}^0$, only the DT events are reconstructed using the missing-mass-squared technique (referred to partially reconstructed events thereafter). 
Therefore, according to Eq.~\ref{eq:NST_PureCP},  the  denominator term in  Eq.~\ref{eq:eq4} can be rewritten as
\begin{equation}
S(f)\epsilon_{\mathrm{DT}}^{g,f}/\epsilon_{\mathrm{ST}}^{f} = 2N_{D\bar{D}}\mathcal{B}(f)\epsilon_{\mathrm{DT}}^{g,f}[1-\eta^{f}_{CP}y],
\label{eq:eq5}
\end{equation}
which can be obtained with the total number of $D^0\bar{D}^0$ pairs quoted from Refs.~\cite{BESIII:NDD1,BESIII:NDD2,BESIII:NDD3}, the branching fraction of the ST mode quoted from  the Particle Data Group (PDG)~\cite{PDG2023}, and the DT efficiency estimated with the signal MC samples. 

For the $CP$-mixed tag modes with full phase space (referred to as global $CP$-mixed hereafter), the expected ST and DT yields are given by
\begin{equation} 
	\label{eq:NST_MixCP}
	S(f)=2N_{D\bar{D}}\mathcal{B}(f)\epsilon_{\mathrm{ST}}^{f}[1-(2F^{f}_{+}-1)y],
\end{equation}
\begin{equation} 
	\label{eq:NDT_MixCP}
	M(g,f)=2N_{D\bar{D}}\mathcal{B}(g)\mathcal{B}(f)\epsilon_{\mathrm{DT}}^{g,f}[1-(2F^{g}_{+}-1)(2F^{f}_{+}-1)].
\end{equation}
\noindent Then $F^{g}_{+}$ can be accessed through
\begin{equation} 
	\label{eq:Fp_MixCP}
	F^{g}_{+}=\frac{N^{+}F^{f}_{+}}{N^{f}-N^{+}+2F^{f}_{+}N^{+}},
\end{equation}
\noindent
with 
\begin{equation} 
N^{f}=\frac{M(g,f)[1-(2F^{f}_{+}-1)y]}{S(f)\epsilon_{\mathrm{DT}}^{g,f}/\epsilon_{\mathrm{ST}}^{f}},
\label{eq:eq9}
 \end{equation}
where $N^{+}$ is taken from the measurements involving pure $CP$ tags, $N^{f}$ is determined with the measured DT and ST yields, the corresponding ST and DT detection efficiencies,  and the $CP$-even fraction of the ST mode. In practice, the $F^{\pi^+\pi^-\pi^+\pi^-}_{+}$ is taken from Ref.~\cite{BESIII:Fp4pi}, and the $F^{\pi^{+}\pi^{-}\pi^{0}}_{+}$ and $F^{K^{+}K^{-}\pi^{0}}_{+}$ are taken from this work. 

For the global $CP$-mixed tag modes which are the same as the signal mode, the expected ST and DT yields are given by
\begin{equation} 
	\label{eq:NST_MixCP_self}
	S(g)=2N_{D\bar{D}}\mathcal{B}(g)\epsilon_{\mathrm{ST}}^{g}[1-(2F^{g}_{+}-1)y]\mathrm{,}
\end{equation}
\begin{equation} 
	\label{eq:NDT_MixCP_self}
	M(g,g)=N_{D\bar{D}}\mathcal{B}^{2}(g)\epsilon_{\mathrm{DT}}^{g,g}[1-(2F^{g}_{+}-1)^{2}]\mathrm{.}
\end{equation}
Then $F^{g}_{+}$ can be accessed through
\begin{equation} 
	\label{eq:Fp_MixCP_self}
	F^{g}_{+}=1-\frac{N^{g}}{2N^{+}}
\end{equation}
\noindent
with
\begin{equation} 
N^{g}=\frac{2M(g,g)[1-(2F^{g}_{+}-1)y]} {S(g)\epsilon_{\mathrm{DT}}^{g,g}/\epsilon_{\mathrm{ST}}^{g}}, 
\label{eq:eq13}
\end{equation}
where $F^{g}_{+}$  is determined with the measured DT and ST yields, the corresponding ST and DT detection efficiencies and $N^+$ are taken from the measurements made with pure $CP$ tags.

For the $CP$-mixed tag modes $D\to K^{0}_{S,L}\pi^{+}\pi^{-}$, the $CP$-even fractions are close to 0.5~\cite{Gershon:Kspipi}. According to Eq.~\ref{eq:NDT_MixCP}, these tag modes integrated over full phase space give very low sensitivity for the $F_{+}^{g}$ the measurement. However, an alternative measurement with different phase-space bins (referred to as binning $CP$-mixed) of the tag modes can be performed. 
Then the DT yields in a specific phase-space bin $i$ for the tag modes $D\to K^{0}_{S,L}\pi^{+}\pi^{-}$ are given by 
\begin{equation} 
	\label{eq:NDT_MixCP_Kspipi}
	M_{i}(g,K^{0}_{S}\pi^{+}\pi^{-})=H[K_{i}+K_{-i}-2\sqrt{K_{i}K_{-i}}c_{i}(2F^{g}_{+}-1)]\mathrm{,}
\end{equation}
\begin{equation} 
	\label{eq:NDT_MixCP_Klpipi}
	M'_{i}(g,K^{0}_{L}\pi^{+}\pi^{-})=H'[K'_{i}+K'_{-i}+2\sqrt{K'_{i}K'_{-i}}c'_{i}(2F^{g}_{+}-1)]\mathrm{.}
\end{equation}
\noindent
where $H(H')$ is a normalization factor, $K_{i}(K'_{i})$ is the fraction of $D^0\to K^{0}_{S}\pi^{+}\pi^{-}(K^{0}_{L}\pi^{+}\pi^{-})$  decays in the $i^{\rm th}$ bin and $c_{i}(c'_{i})$ is the corresponding amplitude-weighted cosine of the average strong-phase difference.
Unlike for the other DTs, here $F^{g}_{+}$ is extracted by constraining the relative partial decay width (proportional to $M_{i}(M'_{i})$) in different phase-space bins, without any need to use the ST yields as input.

\section{BESIII DETECTOR AND MONTE CARLO SIMULATION}

The BESIII detector~\cite{Ablikim:2009aa} records the data from the symmetric $e^{+}e^{-}$ collisions provided by the BEPCII storage ring~\cite{Ring:2016} in the center-of-mass energy ranging from 1.84 to 4.95 GeV, with a peak luminosity of ${1.1\times 10^{33}~\mathrm{cm}^{-2}\mathrm{s}^{-1}}$ achieved at 3.773 GeV. BESIII has collected large data samples in this energy region~\cite{BESIII2020a,EcmsMea,EventFilter}. The cylindrical core of the BESIII detector covers 93\% of the full solid angle and consists of a helium-based multilayer drift chamber (MDC), a plastic scintillator time-of-flight system (TOF), and a CsI(Tl) electromagnetic calorimeter (EMC). All these detector are enclosed in a superconducting solenoidal magnet providing a 1.0 T magnetic field.  The solenoid is supported by an octagonal flux-return yoke with resistive plate counter (RPC) muon-identification (MUC) modules interleaved with steel. The main function of the MUC is to separate muons from charged pions and other hadrons based on their hit patterns in the flux return yoke. The MDC provides the charged-particle momentum with a resolution of 0.5\% at 1 $\mathrm{GeV}/c$, and the energy loss ($\mathrm{d}E/\mathrm{d}x$) resolution of 6\% for electrons from Bhabha scattering. The EMC measures photon with an energy resolution of 2.5\% (5\%) at 1 GeV in the barrel (end-cap) region. The time resolution in the TOF barrel region is 68 ps, while that in the end-cap region was 110 ps. The end-cap TOF system was upgraded in 2015 using multigap resistive plate chamber technology, providing a time resolution of 60 ps, which benefits 63\% of the data used in this analysis~\cite{etofa,etofb,etofc}.

The MC simulated data samples produced with a {\footnotesize{GEANT}}4-based~\cite{Geant4:2002hh} package, which includes the geometric description of the BESIII detector and the detector response, are used to determine detection efficiencies and to estimate backgrounds. In the simulation, the beam-energy spread is implemented and initial-state radiation (ISR) in the $e^{+}e^{-}$ annihilations is modeled with the generator {\footnotesize{KKMC}}~\cite{KKMC:2000,KKMC:2001}. The inclusive MC sample consisting of the production of $D\bar{D}$ and $D^+D^-$ pairs (including quantum coherence for the neutral $D$ channels), the non-$D\bar{D}$ and $D^+D^-$ decays of the $\psi(3770)$, the ISR production of the $J/\psi$ and $\psi(3686)$ states, and the continuum processes are produced by incorporating the {\footnotesize{KKMC}}~\cite{KKMC:2000}. All particle decays are modeled with {\footnotesize{EVTGEN}}~\cite{Lange:2001uf,Ping:2008zz} using branching fractions either taken from the PDG~\cite{PDG2023}, when available, or otherwise estimated with {\footnotesize{LUNDCHARM}}~\cite{Chen:2000tv,Ping:2014}. Final-state radiation from charged final-state particles is incorporated using the {\footnotesize{PHOTOS}} package~\cite{photos2}. 

Signal MC samples for different ST modes are generated individually. In the MC generation, the decays $D\to\pi^{+}\pi^{-}\pi^{0}$ and $D\to K^{+}K^{-}\pi^{0}$ follow an isobar-based amplitude model obtained by fitting the BESIII data sample, where the flavor of the signal charmed meson is inferred by reconstructing the accompanied charmed meson in the event through its decay into a flavor-specific final state.  
The simulated DT samples involving $D\to K^{0}_{S,L}\pi^{+}\pi^{-}$ tag modes are implemented with an amplitude model developed by the BaBar Collaboration~\cite{KslpipiModel:2010}. In generating all the DT samples, quantum correlation effects have been considered to better estimate the reconstruction efficiencies, especially for different phase-space bins of the $D\to K^{0}_{S,L}\pi^{+}\pi^{-}$ tag modes.

\section{EVENT SELECTION}

All the tag modes employed in this analysis are summarized in Table~\ref{tab:tagmode}, where the intermediate states are reconstructed with the following decays: $K^{0}_{S}\to \pi^{+}\pi^{-}$, $\pi^{0}\to\gamma\gamma$, $\eta\to \gamma\gamma$ and $\pi^{+}\pi^{-}\pi^{0}$, $\omega\to\pi^{+}\pi^{-}\pi^{0}$, and $\eta'\to\pi^{+}\pi^{-}\eta$ and $\gamma\rho^{0}$ ($\rho^{0}\to\pi^{+}\pi^{-}$).
DT events with tag modes involving a $K^{0}_{L}$ candidate are partially reconstructed using the missing-mass-squared technique, while the other DT events are fully reconstructed. 

Charged tracks are required to satisfy $|\mathrm{cos}\theta|<0.93$, where $\theta$ is the polar angle in the MDC defined with respect to the $z$-axis, which is the symmetry axis of the MDC. 
For charged tracks not originated from $K^{0}_{S}$ decay, the distance of closest approach to the interaction point (IP) must be less than 10~cm along the $z$-axis ($|V_{z}|$) and less than 1~cm in the transverse plane ($V_{xy}$). Particle identification (PID) for charged tracks is performed by combining information from the $\mathrm{d}E/\mathrm{d}x$ in the MDC and the flight time in the TOF to form likelihoods $\mathcal{L}(h)$  for each hadron $h$ ($K$ and $\pi$) hypothesis, individually. Charged kaons and pions are identified by comparing the obtained likelihoods and requiring $\mathcal{L}(K)>\mathcal{L}(\pi)$ and $\mathcal{L}(\pi)>\mathcal{L}(K)$, respectively

Photon candidates are identified using showers in the EMC. The deposited energy of each shower must be greater than 25 MeV in the barrel region ($|\mathrm{cos}\theta|<0.80$) or 50 MeV in the end-cap region ($0.86<|\mathrm{cos}\theta|<0.92$). To exclude showers originated from charged tracks, the opening angle between the EMC shower and the closest charged track at the EMC must be greater than $10^{\circ}$ as measured from the IP. To suppress electronic noise and showers unrelated to the event, the difference between the EMC time and the event start time must be within $[0,700]$~ns.

The $K^{0}_{S}$ candidates are reconstructed with two oppositely charged tracks satisfying $|V_{z}|<20~\mathrm{cm}$ and without PID imposed. 
The primary vertex and secondary vertex fits are performed on the two charged tracks under the  $\pi^{+}\pi^{-}$ hypothesis to determine the invariant mass and the decay length of  $K^{0}_{S}$ candidates.  
The $K^{0}_{S}$ candidates are required to have a $\pi^{+}\pi^{-}$ invariant mass lying within  $[0.487,0.511]~\mathrm{GeV}/c^{2}$, and a decay length larger than twice the resolution on this distance. The kinematic variables of $K^{0}_{S}$ updated by the primary vertex fit are utilized in the subsequent analysis. 

The $\pi^{0}$ and $\eta$ candidates are reconstructed using photon pairs with $\gamma\gamma$ invariant mass within $[0.115,0.150]$ and $[0.505,0.575]$~$\mathrm{GeV}/c^{2}$, respectively. No candidates are accepted where   both photons fall within the end-cap EMC region. 
In order to improve the momentum resolution, a kinematic fit is carried out on the $\gamma\gamma$ candidates in which the invariant mass of the photon pair is constrained to the known $\pi^{0}$ or $\eta$ mass~\cite{PDG2023}. The kinematic variables from this fit are used in the subsequent analysis.  The $\eta$ candidates are also reconstructed with the decay mode  $\eta\to \pi^{+}\pi^{-}\pi^{0}$, where the invariant mass of the final state is required to lie within  $[0.530,0.565]~\mathrm{GeV}/c^{2}$. Similarly,  $\omega$ candidates are reconstructed with the dominant decay mode of $\omega\to\pi^{+}\pi^{-}\pi^{0}$, with the $\pi^{+}\pi^{-}\pi^{0}$ invariant mass lying within $[0.750,0.820]~\mathrm{GeV}/c^{2}$.
The $\rho^{0}$ candidates are reconstructed with the decay mode of $\rho^{0}\to \pi^{+}\pi^{-}$, with the invariant mass of the $\pi^{+}\pi^{-}$ pair required to lie within $[0.626,0.924]~\mathrm{GeV}/c^{2}$.
The $\eta'$ candidates are reconstructed with the decay mode of $\eta'\to\pi^{+}\pi^{-}\eta(\gamma\gamma)$ and $\eta'\to\gamma\rho^{0}(\pi^+\pi^-)$, with the invariant masses of the $\pi^{+}\pi^{-}\eta$ ($\eta\to\gamma\gamma$) and $\gamma\pi^{+}\pi^{-}$ combinations within the ranges $[0.940,0.976]$ and $[0.940,0.970]~\mathrm{GeV}/c^{2}$, respectively.

ST samples from the 12 tag modes not involving a $K^0_L$ are selected by using the  $\pi^{\pm}$, $K^{\pm}$, $\pi^0$, $K^{0}_{S}$, $\eta$, $\omega$ and $\eta'$ candidates to reconstruct each tag-mode decay, taking account of all  possible  combinations and ensuring that no  charged tracks and photons are used for more than one intermediate state.
To further improve significance, additional requirements are implemented to reduce the backgrounds in some tag modes.
For the modes  $D\to\pi^{+}\pi^{-}\pi^{0}$ and $D\to 2(\pi^{+}\pi^{-})$, candidates are rejected if any ${\pi^{+}\pi^{-}}$ combination has an invariant mass lying in $[0.481,0.514]~\mathrm{GeV}/c^{2}$ to suppress the backgrounds associated with $K^{0}_{S}$ decays.
For the modes $D\to K^{+}K^{-}$ and $D\to \pi^{+}\pi^{-}$, it is necessary to reject contamination from cosmic ray, di-muon and Bhabha events. 
To remove the cosmic-ray background, the two charged tracks are required to have a flight time difference measured with the TOF less than ${\mathrm{5~ns}}$.
To reject the dimuon and Bhabha backgrounds, PID algorithms are deployed  based on the $\mathrm{d}E/\mathrm{d}x$ in the MDC, the flight time in the TOF, the deposited energy in the EMC, and the penetration depth in the MUC for the two charged tracks~\cite{VetoRay}.  
The candidates are rejected if both charged tracks are identified as an $e^{+}e^{-}$ or $\mu^+\mu^-$ pair.
To further suppress background, the tag modes $D\to K^{+}K^{-}$ and $D\to \pi^{+}\pi^{-}$ are further required to have at least one additional shower in the EMC with energy greater than 50~MeV or at least one additional charged track in the MDC.

If multiple combinations exist for a specific mode, the one with the  minimum $|\Delta E|=|E_{D}-\sqrt{s}/2|$ is retained for the subsequent analysis, where $E_{D}$ is the energy of ST $\bar{D}$ candidate obtained by summing  energies of its daughter particles in the center-of-mass frame, and $\sqrt{s}$ is the center-of-mass energy. 
To suppress combinatorial background, the $\Delta E $ for each mode is required to be within $\pm3$ times of its resolution around its peak.

\section{DETERMINATION OF THE ST YIELDS}

\begin{figure*}[hbtp]
	\begin{center}
		\subfigure{\includegraphics[width=0.9\textwidth,height=0.4\textheight]{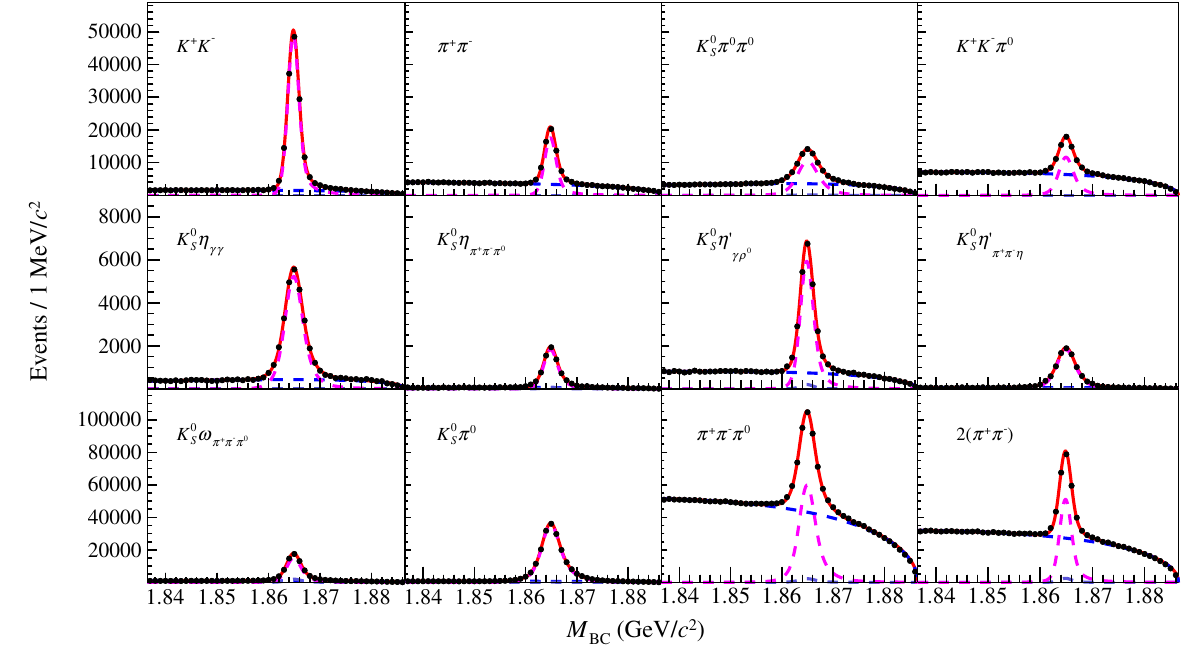}}
	\end{center}
	\setlength{\abovecaptionskip}{-0.2cm}
	\caption{
		Fits to the $M_{\mathrm{BC}}$ distributions of different ST modes. The black dots with error bars are data. The red solid curves indicate the fit results and the blue dashed curves describe the background shapes. The pink and purple dashed curves are the signal and peaking background shapes, respectively.
	} \label{fig:ST_Yield}
\end{figure*}

The ST yields are extracted by fitting the distribution of the beam-constrained mass, defined as ${M_{\mathrm{BC}}=\sqrt{(\sqrt{s}/2)^{2}-|\vec{p}_{\bar{D}}|^{2}}}$,  where $\vec{p}_{\bar{D}}$ is the momentum vector of the ST $\bar{D}^0$ candidate, obtained by vector-summing the momenta of its daughter particles in the center-of-mass frame. Unbinned maximum-likelihood fits are performed on the $\mathrm{M_{\mathrm{BC}}}$ distributions, as shown in Fig.~\ref{fig:ST_Yield} where the signal is described by the simulated shape convolved with a Gaussian function which accounts for the resolution difference between data and MC simulation, and the background component is modeled with an ARGUS function~\cite{ARGUS:1990hfq}, in which the slope parameters are free parameters, and the end point is fixed to the beam energy. 

Studies performed using the inclusive MC sample indicate that there are small peaking backgrounds in the $\mathrm{M_{\mathrm{BC}}}$ distributions from decay modes of the same or similar topologies. 
The dominant backgrounds in the modes ${D\to K^{0}_{S}\pi^{0}~(K^{0}_{S}\pi^{0}\pi^{0})}$ originate from the decays ${D\to \pi^{+}\pi^{-}\pi^{0}~(\pi^{+}\pi^{-}\pi^{0}\pi^{0})}$.
The contamination in the ${D\to K^{0}_{S}\eta'_{\gamma\rho^{0}}~(K^{0}_{S}\eta'_{\pi^{+}\pi^{-}\eta})}$ sample arises from ${D\to K^{0}_{S}\pi^{+}\pi^{-}\pi^{0}}$ and ${D\to K^{0}_{S}\pi^{+}\pi^{-}~(K^{0}_{S}\pi^{+}\pi^{-}\eta)}$ decays. The contamination in the  ${D\to \pi^{+}\pi^{-}\pi^{0}}$ and ${D\to2(\pi^{+}\pi^{-})}$ samples comes  mainly from ${D\to K^{0}_{S}\pi^{0}}$ and ${D\to K^{0}_{S}\pi^{+}\pi^{-}}$, and that for 
 ${D\to K^{0}_{S}\omega}$ and ${K^{0}_{S}\eta_{\pi^{+}\pi^{-}\pi^{0}}}$  arises from ${D\to K^{0}_{S}\pi^{+}\pi^{-}\pi^{0}}$ decays.
The yields of the background events of ${D\to K^{0}_{S}\omega}$ and ${K^{0}_{S}\eta_{\pi^{+}\pi^{-}\pi^{0}}}$  are determined by fitting the $\mathrm{M_{\mathrm{BC}}}$ distributions in data for events with the ${\pi^{+}\pi^{-}\pi^{0}}$ invariant mass lying  in the $\omega$ or $\eta$ sideband regions, while others are estimated with the MC simulation.

The $\Delta E$ requirements, the peaking background fractions ($F_{\mathrm{bkg}}$), the ST yields after background subtraction ($N_{\mathrm{ST}}$) and ST efficiencies ($\epsilon_{\mathrm{ST}}$) for individual ST modes are summarized in Table~\ref{tab:ST}. 

\begin{table}[htbp]
	\centering
	\caption{Summary of the $\Delta E$ requirements, peaking background fractions $F_{\mathrm{bkg}}$, ST yields $N_{\mathrm{ST}}$ and ST efficiencies $\epsilon_{\mathrm{ST}}$ for the ST samples. The uncertainties are statistical only.}	
	\label{tab:ST}
	
	\begin{tabular}[b]{ l  c c  p{1cm}<{\raggedleft} @{ $\pm$ } p{0.5cm}<{\raggedright} c }
		\hline
		\hline
		\small{Mode} & \makebox[0.100\textwidth][c]{\small{$\Delta E~(\mathrm{MeV})$}} & $F_{\mathrm{bkg}}~(\%)$ & \multicolumn{2}{c}{$N_{\mathrm{ST}}$} & \small{$\epsilon_{\mathrm{ST}}~(\%)$} \\
		\hline
		$K^{+}K^{-}$                                                   & $[-21, 20]$ &  ... & 151106 & 423 & 62.8 \\
		$\pi^{+}\pi^{-}$                                                & $[-36, 35]$ &  ... &   56697 & 310 & 68.4 \\
		$K^{0}_{S}\pi^{0}\pi^{0}$                                & $[-72, 53]$ &  1.5  &   60466 & 378 & 15.3 \\  \hline
		$K^{0}_{S}\pi^{0}$                                          & $[-71, 51]$ &  0.4  & 191277 & 489 & 39.8 \\
		$K^{0}_{S}\eta_{\gamma\gamma}$               & $[-38, 36]$ &  ... &   25947 & 189 & 33.1 \\
		$K^{0}_{S}\eta_{\pi^{+}\pi^{-}\pi^{0}}$            & $[-35, 28]$ &  3.9  &     7112  & 100 & 16.4 \\
		$K^{0}_{S}\eta'_{\gamma\rho^{0}}$               & $[-31, 25]$ &  3.3  &   21554  & 182 & 18.6 \\
		$K^{0}_{S}\eta'_{\pi\pi\eta}$                           & $[-35, 31]$ &  1.7  &    9380  &  106 & 14.9 \\
		$K^{0}_{S}\omega_{\pi^{+}\pi^{-}\pi^{0}}$      & $[-42, 33]$ & 11.7 &   61565  &  315 & 15.3 \\ \hline
		$K^{+}K^{-}\pi^{0}$                                         & $[-40, 20]$ & ... &   48332  &  399 & 25.8 \\
		$\pi^{+}\pi^{-}\pi^{0}$                                      & $[-62, 51]$ &  3.7  & 285480 & 1090 & 38.7 \\
		$2(\pi^{+}\pi^{-})$                             & $[-26, 23]$ &  4.7 &  171104 &   695 & 42.1 \\
		\hline	
		\hline
	\end{tabular}
\end{table}

\section{DETERMINATION OF THE DT YIELDS}

The DT candidates for the 12 fully reconstructed tag modes are selected by reconstructing the signal decays  $D\to\pi^{+}\pi^{-}\pi^{0}$ or $D\to K^{+}K^{-}\pi^{0}$ from the remaining charged tracks and reconstructed $\pi^{0}$ candidates against the reconstructed ST $\bar{D}$ candidates. 
Only events with two extra charged tracks and a net charge of zero are considered. Both charged tracks must be identified as pions or kaons. In the signal mode  $D\to\pi^{+}\pi^{-}\pi^{0}$, a $K^{0}_{S}$ veto is applied by imposing the same requirement as that in the ST selection to suppress the background of $D\to K^{0}_{S}\pi^{0}$ decays. If multiple combinations for each signal decay exist, the one with the minimum $|\Delta E|$  is retained for the subsequent analysis. The  $\Delta E$ values must fulfill the same requirements as in the ST selection to suppress the combinatorial background.

\begin{figure*}[htpb]
	\begin{center}
		\subfigure{\includegraphics[width=0.9\textwidth,height=0.4\textheight]{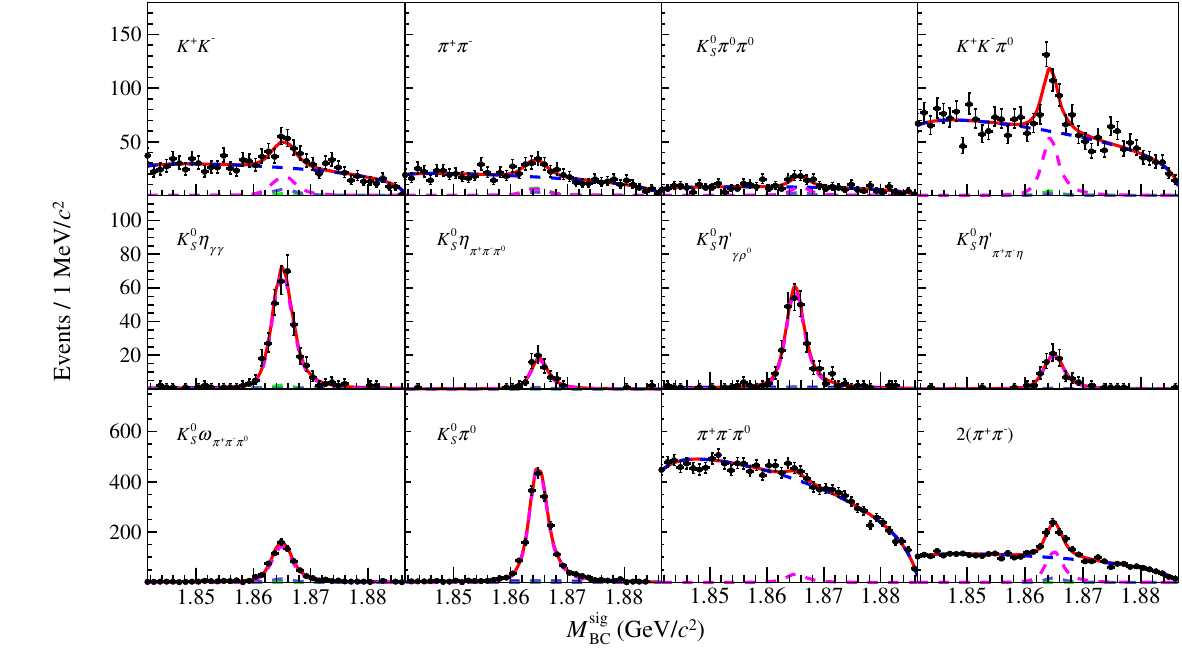}}
		
	\end{center}
     \vskip -0.3cm
	\setlength{\abovecaptionskip}{-0.0cm}
	\caption{
		The projections of the 2D fits on the $M^{\mathrm{sig}}_{\mathrm{BC}}$ distribution for the decay $D\to \pi^{+}\pi^{-}\pi^{0}$. The black dots with error bars are data. The red solid curves represent the fit results and the blue dashed curves describe the non-peaking background shapes. The pink, green and purple dashed curves are the shapes of the signals, the backgrounds with correctly reconstructed signal mode but incorrectly reconstructed tag modes and the very small peaking backgrounds, respectively.  
	} \label{fig:DT_Yield_pipipi0}
\end{figure*}

\begin{figure*}[!htbp]
	\begin{center}
		\subfigure{\includegraphics[width=0.9\textwidth,height=0.4\textheight]{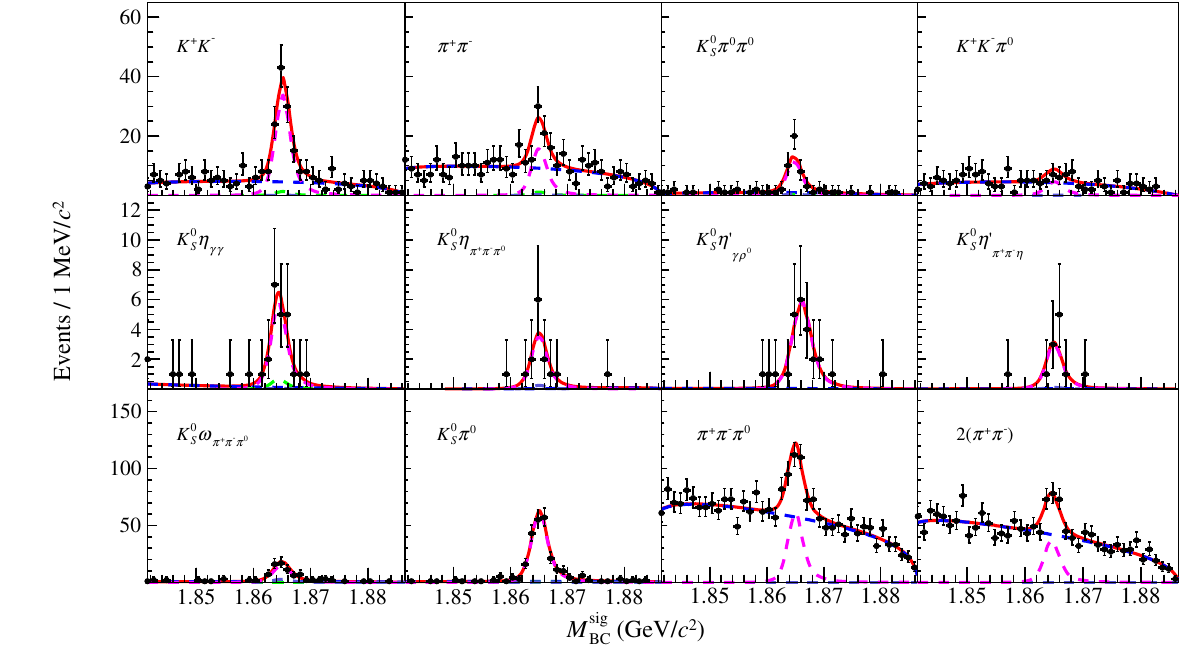}}
		
	\end{center}
      \vskip -0.2cm
	\setlength{\abovecaptionskip}{-0.0cm}
	\caption{
		The projections of the 2D fits on the $M^{\mathrm{sig}}_{\mathrm{BC}}$ distribution for the decay $D\to K^{+}K^{-}\pi^{0}$. The black dots with error bars are data. The red solid curves represent the fit results and the blue dashed curves describe the non-peaking background shapes. The pink, green and purple dashed curves are the shapes of the signals, the backgrounds with correctly reconstructed signal mode but incorrectly reconstructed tag modes and the very small peaking backgrounds, respectively.  
	} \label{fig:DT_Yield_kkpi0}
\end{figure*}

Two-dimensional (2D) unbinned maximum likelihood fits are performed on the distributions of  $M^{\mathrm{tag}}_{\mathrm{BC}}$ versus $M^{\mathrm{sig}}_{\mathrm{BC}}$~\cite{D2Fita,D2Fitb} to determine the DT yields for each tag mode. The signal is described by a 2D simulated shape convolved with the Gaussian resolution functions in each  dimension to account for the  difference in resolution between data and MC simulation, where the means and widths of the Gaussian resolution functions  are obtained from one dimensional fits to the $M^{\mathrm{tag}}_{\mathrm{BC}}$  and $M^{\mathrm{sig}}_{\mathrm{BC}}$ distributions for each tag mode. 
The backgrounds are separated into two categories.  The first comprises events with correctly reconstructed signal mode and incorrectly reconstructed tag mode (or vice versa). This is modeled by the product of the projection of the simulated shape convolved with a Gaussian resolution function for the correctly reconstructed decay and an ARGUS function for the incorrectly reconstructed decay.  The second contains events where neither the signal mode nor tag mode is correctly reconstructed. This is modeled by the product of a student function  in the diagonal band~\cite{D2Fita}  and an ARGUS function for each decay dimension ($M^{\mathrm{tag}}_{\mathrm{BC}}$ and $M^{\mathrm{sig}}_{\mathrm{BC}}$). 
The end point of the ARGUS function is fixed to the beam energy and the others are free parameters in the fit. 
Figures~\ref{fig:DT_Yield_pipipi0} and~\ref{fig:DT_Yield_kkpi0} illustrate the projections of the 2D fits on the $M^{\mathrm{sig}}_{\mathrm{BC}}$ distributions. 

Studies of inclusive MC are used to determine the level and source of the peaking backgrounds in the sample.   The most significant contribution comes from $D\to K^{0}_{S}\pi^{0}$ decays in the $D\to \pi^{+}\pi^{-}\pi^{0}$ sample. Correction factors are applied to account for quantum correlations, according to   Eq.~\ref{eq:NDT_MixCP}. The values of  $F_{+}^{\pi^{+}\pi^{-}\pi^{0}}$ and $F_{+}^{K^{+}K^{-}\pi^{0}}$ are taken from the current analysis, after iterating the estimates until convergence occurs.  The DT yields after background subtraction and the DT detection efficiencies obtained from the corresponding signal MC samples are summarized in Table~\ref{tab:NDT}.

\begin{table}[!htbp]
	\centering
	\caption{Summary of the DT yields $N_{\mathrm{DT}}$ and DT efficiencies $\epsilon_{\mathrm{DT}}~(\%)$ for individual tag modes. The uncertainties on $N_{\rm DT}$ are statistical only.}	
	\label{tab:NDT}
	
	\begin{tabular}[b]{ l p{0.7cm}<{\raggedleft} @{ $\pm$ } p{0.7cm}<{\raggedright} c p{0.7cm}<{\raggedleft} @{ $\pm$ } p{0.7cm}<{\raggedright} c }
		\hline
		\hline
	    \small{Mode} & \multicolumn{2}{c}{$N_{\mathrm{DT}}^{\pi^{+}\pi^{-}\pi^{0}}$} & \small{$\epsilon_{\mathrm{DT}}^{\pi^{+}\pi^{-}\pi^{0}}$} & \multicolumn{2}{c}{$N_{\mathrm{DT}}^{K^{+}K^{-}\pi^{0}}$} & \small{$\epsilon_{\mathrm{DT}}^{K^{+}K^{-}\pi^{0}}$} \\
	    \hline
		$K^{+}K^{-}$ &  91 & 13 & 27.3 & 114 & 13 & 19.5 \\
		
		$\pi^{+}\pi^{-}$ &  31 & 8 & 29.1 & 55 & 11 & 21.4 \\
		
		$K^{0}_{S}\pi^{0}\pi^{0}$ &  34 & 8 & 6.2 & 39 & 7 & 4.4 \\
		
		$K^{0}_{L}\omega$ & 86 & 15 & 7.6 & 71 & 12 & 5.3 \\
		
		$K^{0}_{L}\pi^{0}$ & 171 & 18 & 18.3 & 165 & 21 & 12.5 \\ \hline
		
		$K^{0}_{S}\pi^{0}$ &  1921 & 47 & 16.0 & 226 & 16 & 11.0 \\
		
		$K^{0}_{S}\eta_{\gamma\gamma}$ &  313 & 19 & 13.9 & 20 & 5 & 9.1 \\
		
		$K^{0}_{S}\eta_{3\pi}$ &  68 & 9 & 6.8 & 13 & 4 & 4.4 \\
		
		$K^{0}_{S}\eta'_{\gamma\rho^{0}}$ &  247 & 17 & 7.4 & 25 & 5 & 4.9 \\
		
		$K^{0}_{S}\eta'_{\pi\pi\eta}$ &  87 & 9 & 5.9 & 11 & 4 & 3.8 \\
		
		$K^{0}_{S}\omega$ &  652 & 29 & 6.3 & 59 & 9 & 4.2 \\
		
		$K^{0}_{L}\pi^{0}\pi^{0}$ & 898 & 105 & 5.2 & 76 & 15 & 3.0 \\ \hline
		
		$K^{+}K^{-}\pi^{0}$ &  205 & 24 & 11.8 & 18 & 6 & 6.9 \\
		
		$\pi^{+}\pi^{-}\pi^{0}$ &  152 & 40 & 15.1 & 210 & 25 & 11.7 \\
		
		$2(\pi^+\pi^-)$ &  479 & 30 & 17.1 & 130 & 19 & 12.7 \\

		\hline	
		\hline
	\end{tabular}
		
\end{table}

\begin{figure*}[!htp]
	\begin{center}
           \subfigure{\includegraphics[width=0.8\textwidth,height=0.28\textheight]{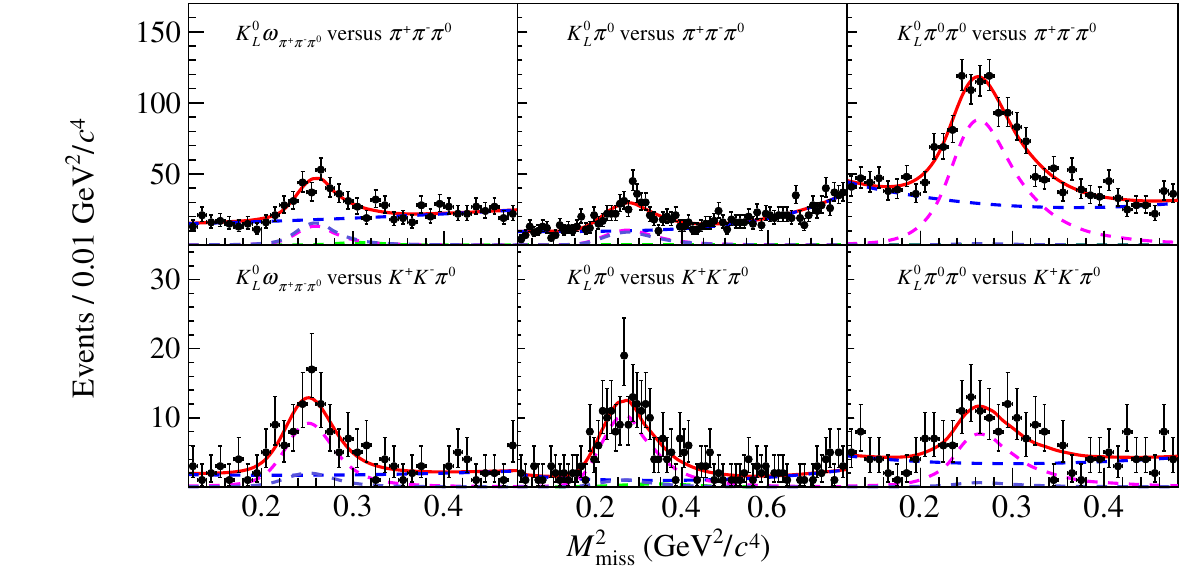}}  
	\end{center}
     \vskip -0.5cm
	\caption{
		Fits to the $M^{2}_\mathrm{miss}$ distributions of the tag modes (Left) $D\to K^{0}_{L}\omega_{\pi^{+}\pi^{-}\pi^{0}}$, (Middle) $D\to K^{0}_{L}\pi^{0}$ and (Right) $D\to K^{0}_{L}\pi^{0}\pi^{0}$. The top row is for the signal decay $D\to \pi^+\pi^-\pi^0$ and the bottom row for the signal decay $D\to K^+K^-\pi^0$.
         The black dots with error bars are data. The red solid curves represent the fit results and the blue dashed curves describe the non-peaking background shapes on the $M^\mathrm{sig}_\mathrm{BC}$ distribution. The pink, green and purple dashed curves show the signal, the $\eta X$ background and the peaking backgrounds, respectively.
	} \label{fig:DT_Yield_Kl}
\end{figure*}

The DT samples involving the tag modes $D\to K^{0}_{L}X$ are partially reconstructed, and the corresponding yields are determined by fitting the missing-mass-squared ($M^{2}_\mathrm{miss}$) distribution.
The signal candidates for $D\to \pi^{+}\pi^{-}\pi^{0}$ or $D\to K^{+}K^{-}\pi^{0}$ are firstly reconstructed with the combination of charged track pairs ($\pi^{+}\pi^{-}$ or $K^{+}K^{-}$) and $\pi^0$.
As in the ST selection, if multiple combinations exist, that one with minimum $|\Delta E|$ is selected.
The $\Delta E$ of each candidate is required to lie within $\pm 3$ times its resolution (which is same as that in the ST selection, as shown in Table~\ref{tab:ST}) and $M_\mathrm{BC}$ $\in$ $[1.86,1.87]$~$\mathrm{GeV}/c^{2}$.
Then, the accompanying particle on the tag side, designated $X$, where $X=\pi^0$,~$\pi^0\pi^0$ or $\omega_{\pi^+\pi^-\pi^0}$,  is selected from the unused neutral and charged pions in the event, where any duplication of the charged tracks and photon  is not allowed.
Any event with any additional charged track, $\pi^{0}$ or $\eta\to\gamma\gamma$ candidate is rejected.
The DT yields are determined from the unbinned maximum-likelihood fits on $M^{2}_\mathrm{miss}$ defined as
\begin{equation} 
	\label{eq:M2_miss}
	M^{2}_\mathrm{miss}=(\sqrt{s}/2-E_{X})^{2}-|\vec{p}_{X}+\hat{p}_{\mathrm{sig}}\sqrt{s/4-{\it{m}}^{2}_{D}}|^{2},
\end{equation}
\noindent
where $E_{X}$ and $\vec{p}_{X}$ are the sum of the reconstructed energies and momentum vectors of $X$, respectively, $\hat{p}_{\mathrm{sig}}$ is the unit momentum vector of $D \to \pi^{+}\pi^{-}\pi^{0}$ or $D\to K^{+}K^{-}\pi^{0}$ signal candidate, and ${\it{m}}^{2}_{D}$ is the known $D^{0}$  mass~\cite{PDG2023}. 
In the fit, the signal is modeled with the MC simulated shape convolved with a Gaussian resolution function with free parameters.  The combinational background is modeled by a second-order Chebychev polynomial function with free parameters. The backgrounds of $D\to K^{0}_{L}\pi^{0}\pi^{0}$, $\pi^{0}\pi^{0}$ and $\eta\pi^{0}$ in the $D\to K^{0}_{L}\pi^{0}$ tag mode, the backgrounds of $D\to \eta\omega_{\pi^{+}\pi^{-}\pi^{0}}$ in the $D\to K^{0}_{L}\omega_{\pi^{+}\pi^{-}\pi^{0}}$ tag mode as well as the backgrounds of $D\to \pi^{0}\pi^{0}\pi^{0}$ and $\eta\pi^{0}\pi^{0}$ in the $D\to K^{0}_{L}\pi^{0}\pi^{0}$ tag mode are modeled by the MC simulated shape with floated yields. The peaking backgrounds are mainly from $D\to K^{0}_{S}\pi^{0}$, $K^{0}_{S}\pi^{0}\pi^{0}$ and $K^{0}_{S}\omega_{\pi^{+}\pi^{-}\pi^{0}}$ for the decays of $D\to K^{0}_{L}\pi^{0}$, $K^{0}_{L}\pi^{0}\pi^{0}$ and $K^{0}_{L}\omega_{\pi^{+}\pi^{-}\pi^{0}}$ in the tag side, respectively.  The signal mode  $D\to\pi^{+}\pi^{-}\pi^{0}$ also has  peaking background from $D\to K^{0}_{S}\pi^{0}$ in the signal side. All these peaking backgrounds are estimated from the inclusive MC sample and corrected for  quantum-correlation effects, where the procedure is iterated until the values of   $F_{+}^{\pi^{+}\pi^{-}\pi^{0}}$ and $F_{+}^{K^{+}K^{-}\pi^{0}}$ converge. 
In the tag mode $D\to K^{0}_{L}\omega_{\pi^{+}\pi^{-}\pi^{0}}$, the non-resonant background, $D\to K^{0}_{L}\pi^{+}\pi^{-}\pi^{0}$, is estimated from the  sideband events in the $\omega$ mass distribution. The fits on the $M^{2}_\mathrm{miss}$ distributions are shown in Fig.~\ref{fig:DT_Yield_Kl} and the corresponding DT yields after background subtraction and the DT efficiencies estimated with signal MC samples are  summarized in Table~\ref{tab:NDT}.

\begin{figure*}[bhtp]
	\begin{center}

            \subfigure{\includegraphics[width=0.9\textwidth,height=0.49\textheight]{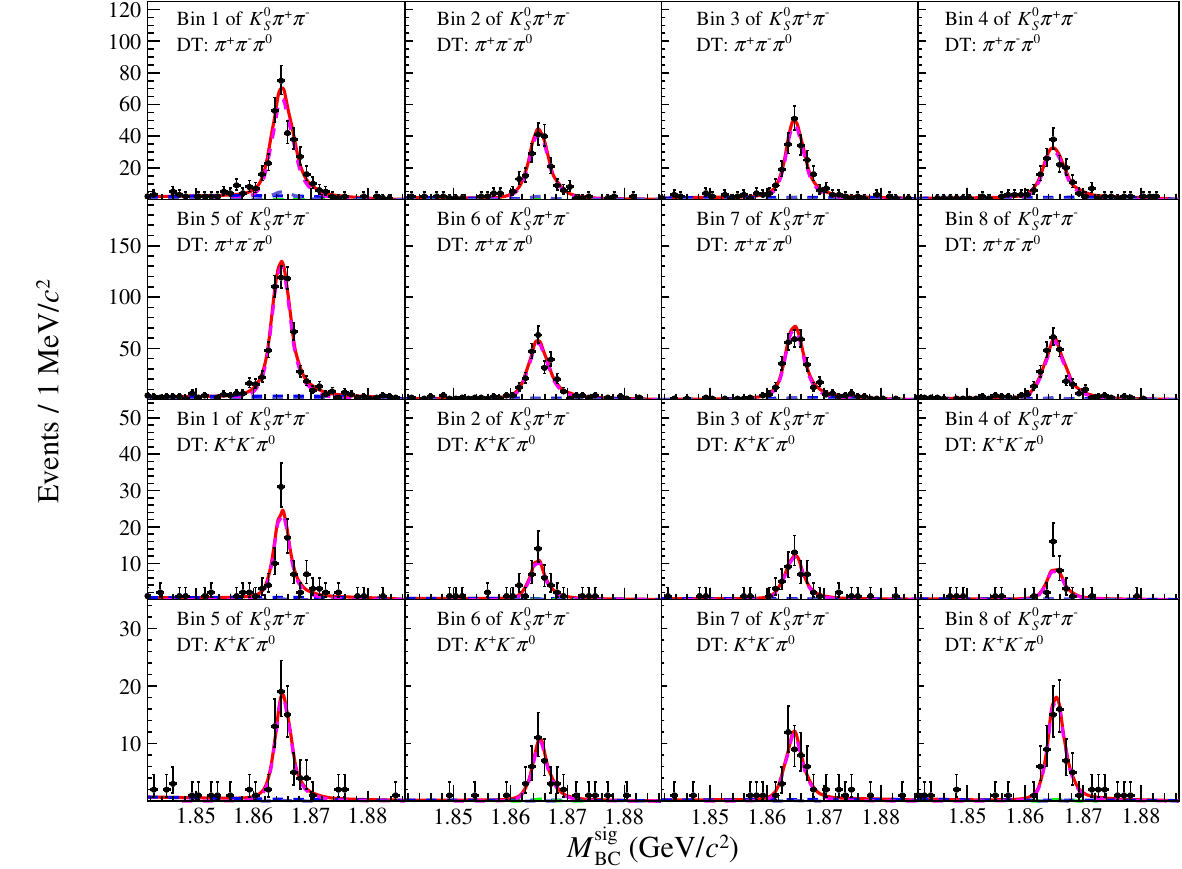}}   
		
	\end{center}
	\setlength{\abovecaptionskip}{-0.2cm}
	\caption{
		The projections of the 2D fits on the $M^\mathrm{sig}_\mathrm{BC}$ distributions for the tag mode $D\to K_{S}^{0}\pi^{+}\pi^{-}$ in individual phase-space bins. The top two rows are for the $D\to\pi^{+}\pi^{-}\pi^{0}$ signal decay, and the bottom two rows for the $D\to K^{+}K^{-}\pi^{0}$ signal decay. The black dots with error bars are data. The red solid curves represent the fit results and the blue dashed curves describe the non-peaking background shapes. The pink, green and purple dashed curves are the shapes of the signals, the backgrounds with correctly reconstructed signal mode but incorrectly reconstructed tag modes and the peaking backgrounds, respectively.
	} \label{fig:DT_Yield_Kspipi}
\end{figure*}

\begin{figure*}[!htp]
	\begin{center}
		
           \subfigure{\includegraphics[width=0.9\textwidth,height=0.49\textheight]{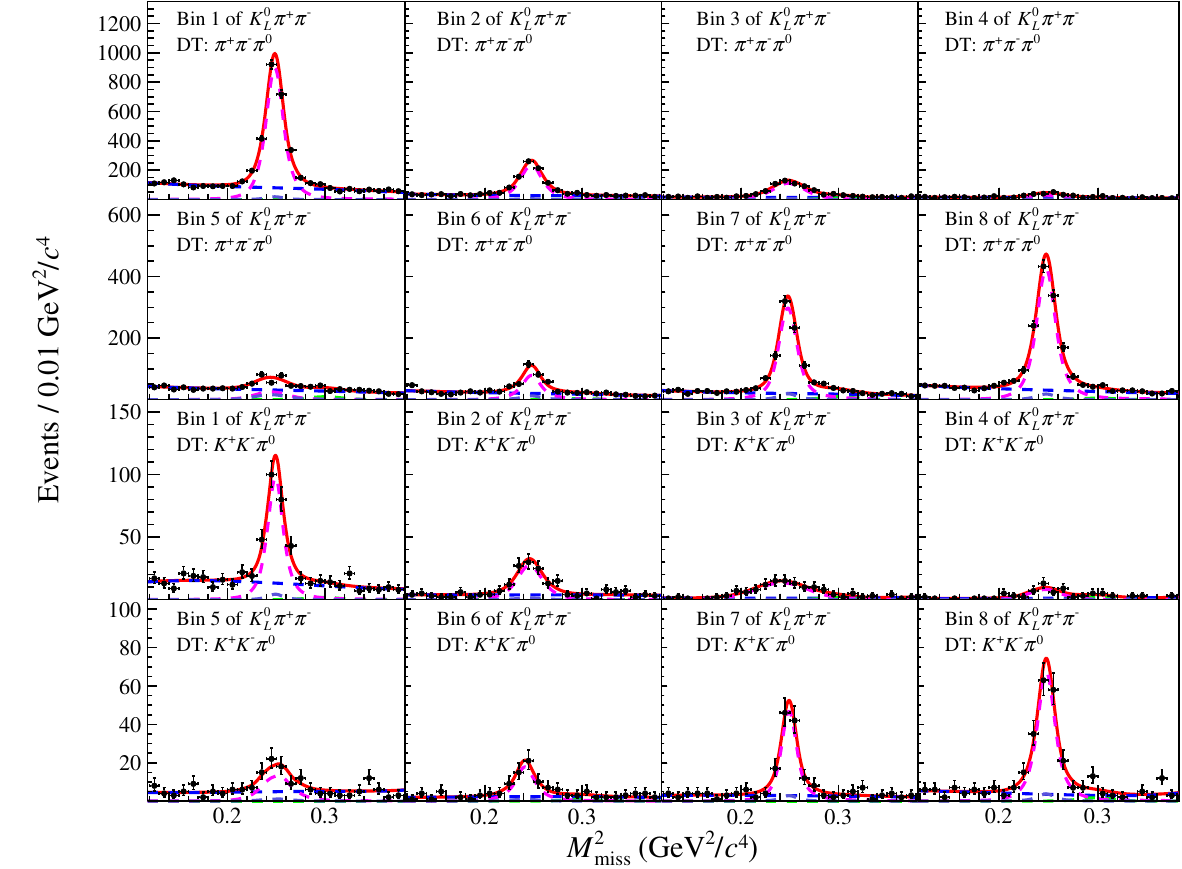}}    
		
	\end{center}
	\setlength{\abovecaptionskip}{-0.2cm}
	\caption{
		Fits to the $M^{2}_\mathrm{miss}$ distributions for the ST mode $D\to K^{0}_{L}\pi^{+}\pi^{-}$ in individual phase-space bins. The top two rows are for the $D\to \pi^{+}\pi^{-}\pi^{0}$ signal decay, and the bottom two rows for the $D\to K^{+}K^{-}\pi^{0}$ signal decay. The black dots with error bars are data. The red solid curves represent the fit results and the blue dashed curves describe the non-peaking background shapes. The pink, green and purple dashed curves are the shapes of the signals, the $\eta\pi^{+}\pi^{-}$ backgrounds and the peaking backgrounds, respectively.
	} \label{fig:DT_Yield_Klpipi}
\end{figure*}

The DT yields for the tag modes ${D\to K^{0}_{S,L}\pi^{+}\pi^{-}}$ are extracted in bins of  phase space, employing  
the ``equal $\Delta \delta$ binning" scheme (8 bins in total), which is used in the CLEO-c and BESIII experiments for the measurement of relative strong-phase difference of this decay~\cite{CLEO:Kslpipi,BESIII:Kslpipi}.
To minimize the migration effect between different bins, additional kinematic fits  are performed in which the invariant mass of  the $K^{0}_{S}\pi^{+}\pi^{-}$ decay is constrained to the known $D^{0}$  mass~\cite{PDG2023} and the mass of the  missing particle in the $K^{0}_{L}\pi^{+}\pi^{-}$ decay is constrained  to the known $K^{0}_{L}$  mass~\cite{PDG2023}. The updated kinematic variables of daughter particles are used to calculate the phase-space bins.
The DT yields in individual phase-space bins are determined with a similar approach as described above, i.e. by performing 2D fits for the tag mode $D\to K^{0}_{S}\pi^{+}\pi^{-}$, and fitting the $M^{2}_\mathrm{miss}$ distribution for the tag mode  $D\to K^{0}_{L}\pi^{+}\pi^{-}$.
In fitting the $M^{2}_\mathrm{miss}$ distribution, the combinatorial background is modeled with a third-order Chebychev polynomial function with free parameters, while the backgrounds from $D\to K^{0}_{L}\pi^{0}\pi^{+}\pi^{-}$, $\eta\pi^{+}\pi^{-}$ in the $D\to K^{0}_{L}\pi^{+}\pi^{-}$ mode are modeled with the MC simulated shape with floated yield. The peaking backgrounds $D\to \pi^{+}\pi^{-}\pi^{+}\pi^{-}$ in the tag mode $D\to K^{0}_{S}\pi^{+}\pi^{-}$ and $D\to K^{0}_{S}\pi^{0}$ in the $D\to \pi^{+}\pi^{-}\pi^{-}$ signal mode are estimated by analyzing the inclusive MC sample and correcting for the quantum-correlation effects.
The projections of the 2D fits on the $M^\mathrm{sig}_\mathrm{BC}$ distributions for the tag mode $D\to K^{0}_{S}\pi^{+}\pi^{-}$ and the $M^{2}_\mathrm{miss}$ distributions for  the tag mode $D\to K^{0}_{L}\pi^{+}\pi^{-}$ in individual phase-space bins are shown in Figs.~\ref{fig:DT_Yield_Kspipi} and \ref{fig:DT_Yield_Klpipi}, respectively. The measured signal yields in individual bins are summarized in Table~\ref{tab:NDT_Kslpipi}.

\begin{table}[!htbp]
	\centering
	\caption{Summary of the DT yields ($N_{\mathrm{DT}}$) for $K^{0}_{S}\pi^{+}\pi^{-}$ and $K^{0}_{L}\pi^{+}\pi^{-}$ in each bin. 
         The uncertainties  are statistical only.}	
	\label{tab:NDT_Kslpipi}
	
	\begin{tabular}[b]{ l p{1.2cm}<{\raggedleft} @{ $\pm$ } p{1.2cm}<{\raggedright} p{1.2cm}<{\raggedleft} @{ $\pm$ } p{0.8cm}<{\raggedright} }
		\hline
		\hline
		\small{$K^{0}_{S}\pi^{+}\pi^{-}$} & \multicolumn{2}{c}{$N_{\mathrm{DT}}^{\pi^{+}\pi^{-}\pi^{0}}$} & \multicolumn{2}{c}{$N_{\mathrm{DT}}^{K^{+}K^{-}\pi^{0}}$} \\
		\hline
		\small{Bin 1} & 291 & 18 & 84 & 10  \\
		
		\small{Bin 2} & 184 & 14 & 36 & 6   \\
		
		\small{Bin 3} & 197 & 15 & 47 & 7   \\
		
		\small{Bin 4} & 144 & 13 & 30 & 6   \\
		
		\small{Bin 5} & 553 & 25 & 65 & 8   \\
		
		\small{Bin 6} & 243 & 17 & 36 & 7   \\
		
		\small{Bin 7} & 291 & 19 & 44 & 7   \\
		
		\small{Bin 8} & 251 & 17 & 65 & 8   \\
		
		\hline
		\small{$K^{0}_{L}\pi^{+}\pi^{-}$} & \multicolumn{2}{c}{$N_{\mathrm{DT}}^{\pi^{+}\pi^{-}\pi^{0}}$} & \multicolumn{2}{c}{$N_{\mathrm{DT}}^{K^{+}K^{-}\pi^{0}}$} \\
		\hline
		\small{Bin 1} & 2407 & 58 & 237 & 19  \\
		
		\small{Bin 2} & 803 & 34 & 103 & 12   \\
		
		\small{Bin 3} & 533 & 29 & 97 & 14   \\
		
		\small{Bin 4} & 146 & 17 & 38 & 8   \\
		
		\small{Bin 5} & 121 & 17 & 56 & 12   \\
		
		\small{Bin 6} & 244 & 21 & 56 & 9   \\
		
		\small{Bin 7} & 847 & 37 & 115 & 12   \\
		
		\small{Bin 8} & 1235 & 40 & 184 & 16   \\
		
		\hline	
		\hline
	\end{tabular}
	
\end{table}

\section{$\textit{\textbf{CP}}$-EVEN FRACTION MEASUREMENT}

\subsection{The $\textit{\textbf{CP}}$-eigen ST modes}
\label{subsec:CPFp}
\begin{figure*}[!htp]
	\begin{center}
		\subfigure[~$N^{+}$ for $\pi^{+}\pi^{-}\pi^{0}$]{\includegraphics[width=0.4\textwidth,height=0.22\textheight]{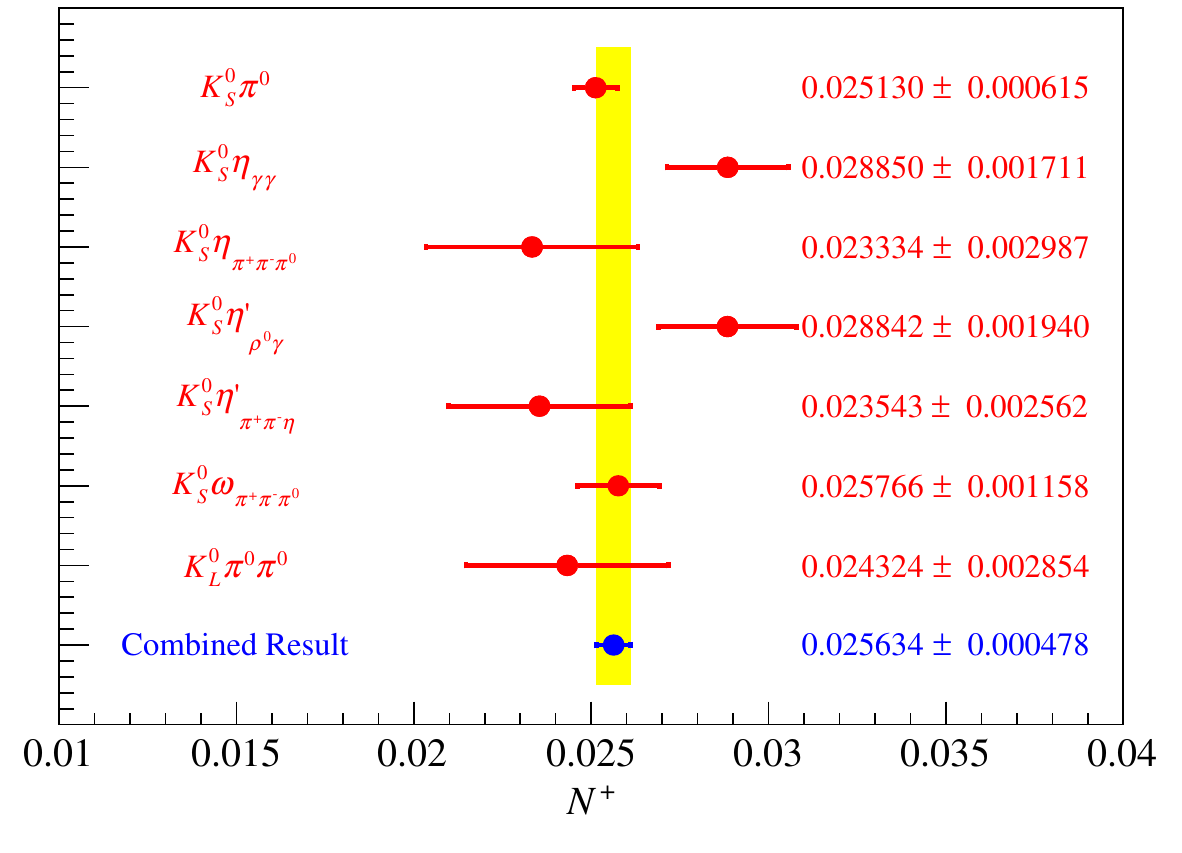}}
		\subfigure[~$N^{-}$ for $\pi^{+}\pi^{-}\pi^{0}$]{\includegraphics[width=0.4\textwidth,height=0.22\textheight]{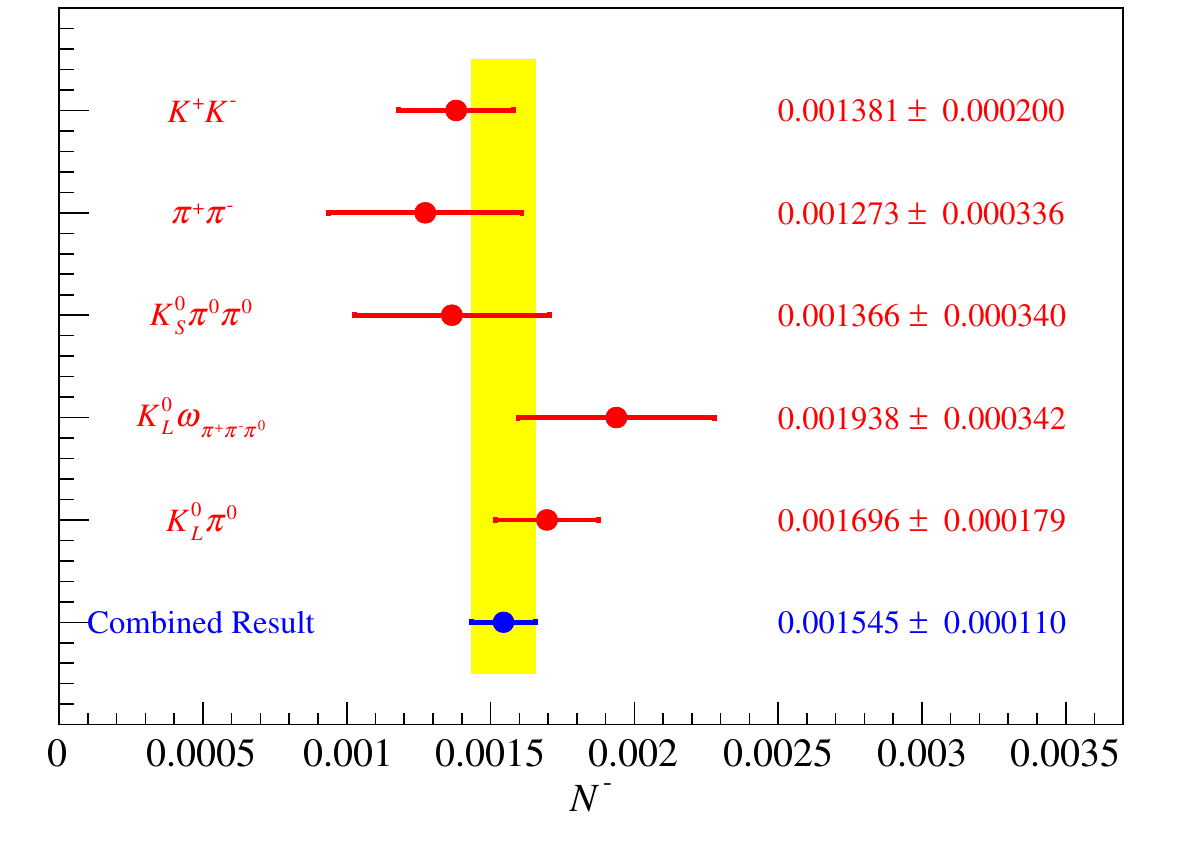}}
			
		\subfigure[~$N^{+}$ for $K^{+}K^{-}\pi^{0}$]{\includegraphics[width=0.4\textwidth,height=0.22\textheight]{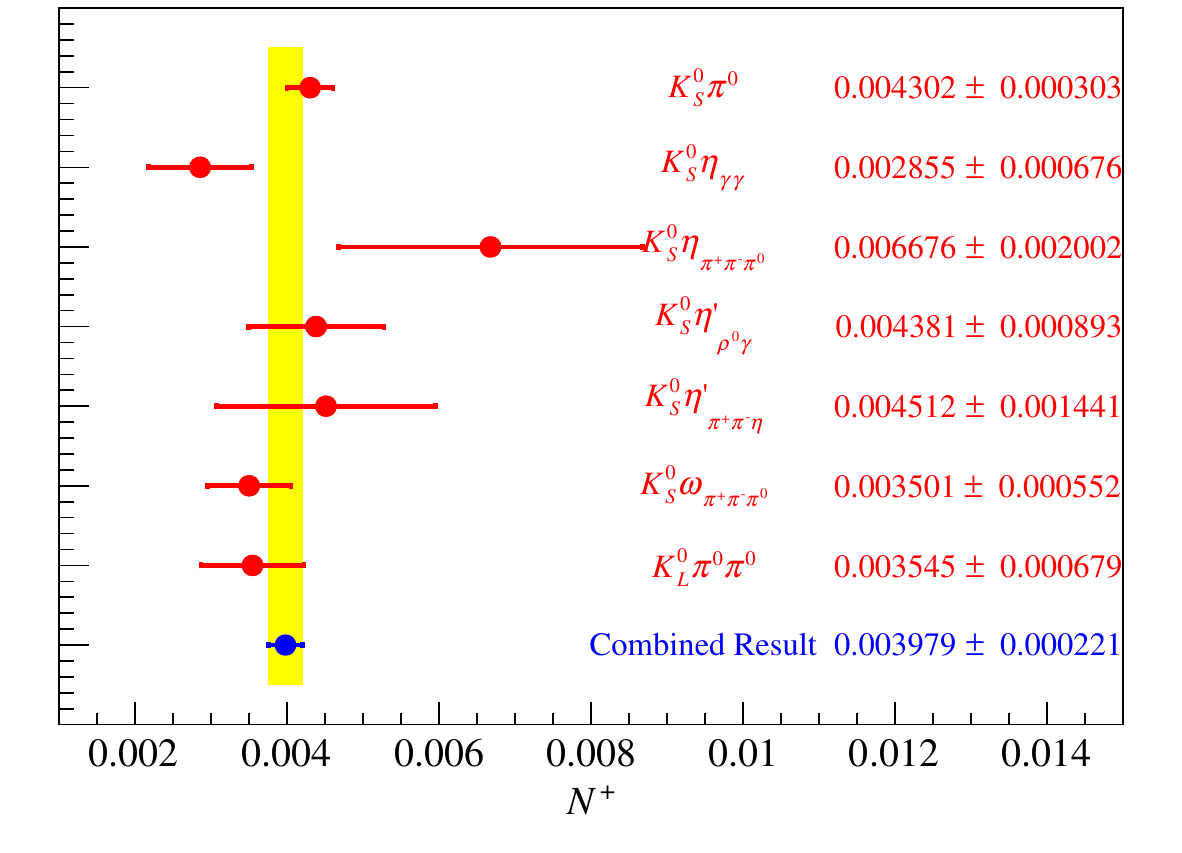}}
		\subfigure[~$N^{-}$ for $K^{+}K^{-}\pi^{0}$]{\includegraphics[width=0.4\textwidth,height=0.22\textheight]{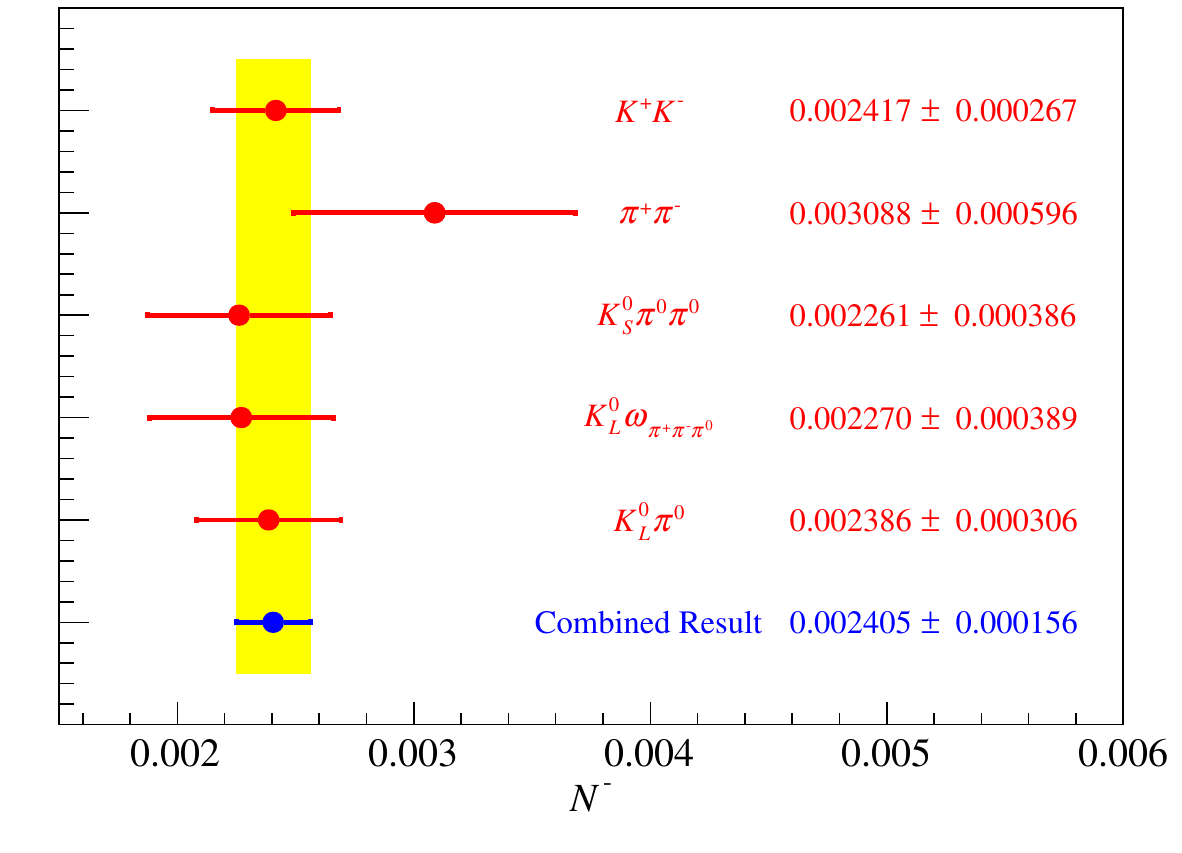}}
		
	\end{center}
	\setlength{\abovecaptionskip}{-0.2cm}
	\caption{
		The individual and average $N^{+}$ and $N^{-}$ for (a, b) $D\to\pi^{+}\pi^{-}\pi^{0}$ and (c,d) $D\to K^{+}K^{-}\pi^{0}$, respectively. 
          The red dots with error bars are for different ST modes with statistical uncertainty. 
          The blue dots with error bars are the average from the least-square fit. 
         The yellow bands correspond to the $\pm 1\sigma$ regions around the average values. 
	} \label{fig:Npm}
\end{figure*}

Using the $CP$-eigen tag modes, the $F_+^{g}$ introduced in Eq.~\ref{eq:Fp_PureCP} is calculated with $N^+$ and $N^-$, which can be extracted according to Eqs.~\ref{eq:eq4} or \ref{eq:eq5} by 
using any $CP$-odd and $CP$-even tag modes, respectively.
To improve the uncertainties, the average  $N^{+}$ and $N^{-}$ for all  $CP$-odd and $CP$-even tag modes shown in Table~\ref{tab:tagmode} are obtained by performing a least-square ($\chi^2$) fit, where the $\chi^2$ is defined as
\begin{equation} 
	\label{eq:chi2_Npm}
	\chi^{2}_{\pm}=\sum_{f}\frac{(N^{\pm}_{f}- \langle N^{\pm}\rangle)^{2}}{\sigma^{2}(N^{\pm}_{f})},
\end{equation}
\noindent
where $\langle N^{\pm} \rangle$ is the expected value of all $CP$-tag modes, and 
$N^{\pm}_{f}$ and $\sigma(N^{\pm}_{f})$ are the corresponding value and uncertainty of the individual contributions.
In the fit, $N^{\pm}_f$ are assumed to be independent and the uncertainty correlation among $D\to K^{0}_{L}X$ tags introduced due to the common input of ${N_{D\bar{D}}=(28,655\pm323)\times 10^{3}}$~\cite{BESIII:NDD1,BESIII:NDD2,BESIII:NDD3} is ignored. The individual $N^{\pm}_{f}$ and the average results are shown in Fig.~\ref{fig:Npm}. 
According to  Eq.~\ref{eq:Fp_PureCP} and the obtained $\langle N^{\pm} \rangle$ for the $CP$-eigen ST modes, we measure  $F_{+}^{\pi^+\pi^-\pi^0}=0.9432\pm0.0040$ and $F_{+}^{K^+K^-\pi^0}=0.623\pm0.020$, where the uncertainties are statistical only. 

\subsection{The global $\textit{\textbf{CP}}$-mixed ST modes}

The $CP$-even fraction $F^g_+$ is measured with the global $CP$-mixed tag modes according to  Eq.~\ref{eq:Fp_MixCP_self} for the self-tag mode (i.e. where the tag and signal modes are the same) and Eq.~\ref{eq:Fp_MixCP}  for the others ($K^{+}K^{-}\pi^{0}$, $\pi^{+}\pi^{-}\pi^{+}\pi^{-}$ for $\pi^{+}\pi^{-}\pi^{0}$ signal mode and $\pi^{+}\pi^{-}\pi^{0}$, $\pi^{+}\pi^{-}\pi^{+}\pi^{-}$ for $K^{+}K^{-}\pi^{0}$ signal mode), where $ N^+$ is taken from the average  $\langle N^+\rangle$ with the $CP$-eigen tag modes,  $N^{f/g}$ is extracted according to Eqs.~\ref{eq:eq9}, and~\ref{eq:eq13} with $CP$-even fraction $F_{+}^{f/g}$, respectively. The measured ST and DT yields and the corresponding detection efficiencies are summarized in Tables~\ref{tab:ST} and \ref{tab:NDT}.
The $N^+$ and $N^{f/g}$, as well as the obtained $F_{+}^g$ in individual global ${CP}$-mixed tag modes are summarized in Table~\ref{tab:fcp-mixing}.
In the calculations, the  $CP$-even fraction $F^{\pi^+\pi^-\pi^+\pi^-}_+=0.735\pm0.015\pm0.005$ is taken from Ref.~\cite{BESIII:Fp4pi}. The values of $F^{\pi^+\pi^-\pi^0}_+$ and  $F^{K^+K^-\pi^0}_+$ are taken from the current analysis, where the calculation is iterated until  convergence is achieved.
 
\begin{table}[!htbp]
	\centering
	\caption{Summary of the $N^+$, $N^{f/g}$ and $F_{+}$ in individual global $CP$-mixed ST modes, where the uncertainties are statistical only.}	
	\label{tab:fcp-mixing}
	
	\begin{tabular}[b]{ l p{4cm} p{4cm} p{4cm}}
		\hline
		\hline
		\small{ST mode} & \multicolumn{1}{c}{~~~$K^{+}K^{-}\pi^{0}$~~~} & \multicolumn{1}{c}{~~~$\pi^{+}\pi^{-}\pi^{0}$~~~}  &  \multicolumn{1}{c}{~~~$\pi^{+}\pi^{-}\pi^{+}\pi^{-}$~~~} \\
		\hline
		
		    &                \multicolumn{3}{c}{$D\to\pi^+\pi^-\pi^0$} \\  \hline
            $N^+$   &   \multicolumn{3}{c}{$0.0256\pm0.0005$}                   \\
            $N^{f/g}$  & \multicolumn{1}{c}{$0.0093\pm0.0011$} & \multicolumn{1}{c}{$0.0027\pm0.0007$} & \multicolumn{1}{c}{$0.0069\pm0.0004$}  \\
             $F_{+}^g$   & \multicolumn{1}{c}{$1.0060\pm0.0675$}  &  \multicolumn{1}{c}{$0.9472\pm0.0139$}  & \multicolumn{1}{c}{$0.9948\pm0.0230$}        \\
           \hline
	        &  \multicolumn{3}{c}{$D\to K^+K^-\pi^0$} \\   \hline
            $N^+$   & \multicolumn{3}{c}{$0.0040\pm 0.0002$}                      \\
            $N^{f/g}$ & \multicolumn{1}{c}{$0.0028\pm0.0010$} & \multicolumn{1}{c}{$0.0024\pm0.0003$} & \multicolumn{1}{c}{$0.0025\pm0.0004$}                   \\
            $F_{+}^g$    & \multicolumn{1}{c}{$0.649\pm0.125$}  &  \multicolumn{1}{c}{$0.631\pm0.030$}  & \multicolumn{1}{c}{$0.667\pm0.058$}       \\
		\hline	
		\hline
	\end{tabular}
	
\end{table}

\subsection{Binning $\textit{\textbf{CP}}$-mixed tag modes}

The measurement of $F^{g}_+$ with the tag modes ${D\to K^{0}_{S,L}\pi^{+}\pi^{-}}$ is performed by analyzing the DT yields in different phase-space bins.
The fractions $K_{i}(K'_{i})$ and amplitude-weighted cosine of the average strong-phase difference $c_{i}(c'_{i})$ in different phase-space bins that enter into Eqs.~\ref{eq:NDT_MixCP_Kspipi} and~\ref{eq:NDT_MixCP_Klpipi}
are taken from Ref.~\cite{BESIII:Kslpipi}.  Migration between different phase-space bins due to the finite kinematic resolution  is parameterized with an efficiency matrix $\epsilon$ from MC simulation, which is defined as
\begin{equation} 
	\label{eq:eff_matrix}
	\epsilon_{ij}=\frac{N^{\mathrm{rec}}_{ij}}{N^{\mathrm{gen}}_{j}},
\end{equation}
\noindent
where $N^{\mathrm{rec}}_{ij}$ is the number of signal MC events produced in the $j^{\rm th}$ phase-space bin but reconstructed in the $i^{\rm th}$ phase-space bin, $N^{\mathrm{gen}}_{j}$ is the number of signal MC events produced in the $j^{\rm th}$ bin. The full efficiency matrices for the $D\to K^{0}_{S,L}\pi^{+}\pi^{-}$  ST modes can be found in Appendix~\ref{app:eff_matrix}.

\begin{figure*}[!htp]
	\begin{center}
		\subfigure[~$K_{S}^{0}\pi^{+}\pi^{-}$ versus $\pi^{+}\pi^{-}\pi^{0}$]{\includegraphics[width=0.4\textwidth,height=0.22\textheight]{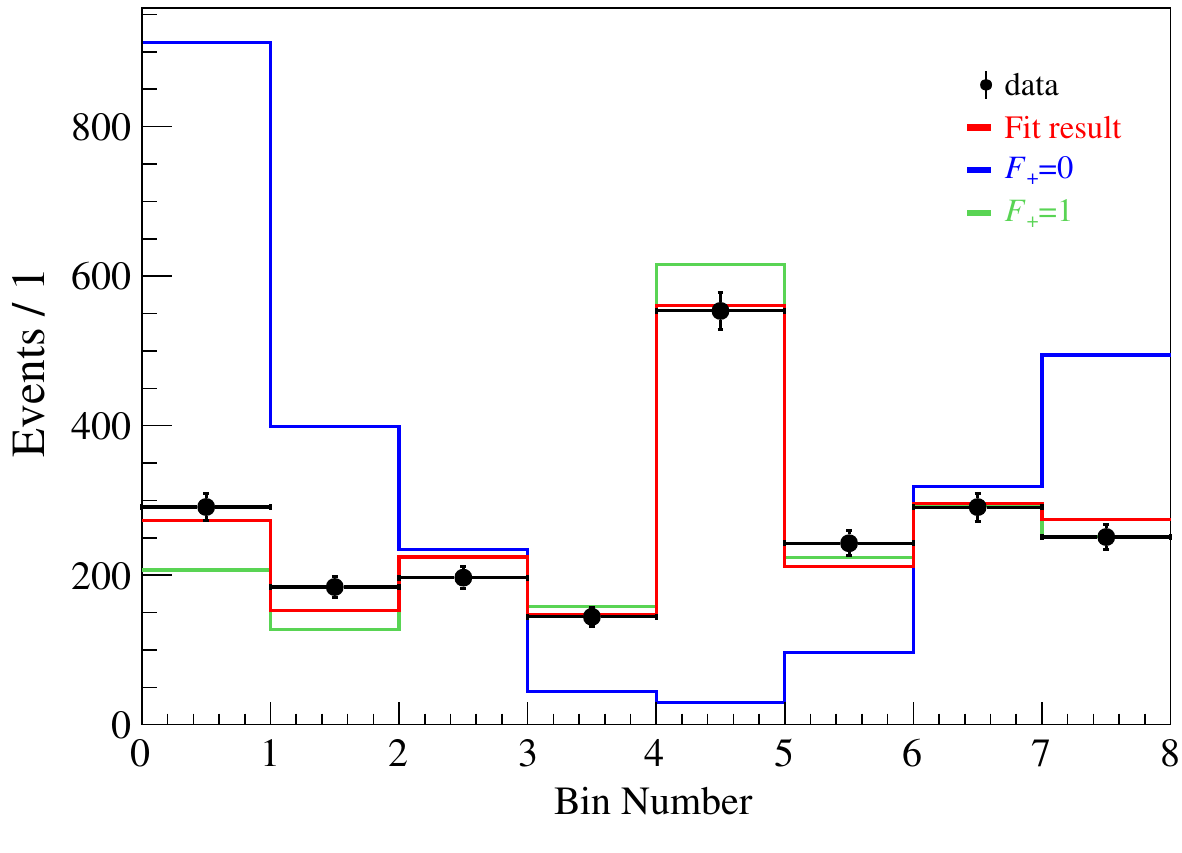}}
		\subfigure[~$K_{L}^{0}\pi^{+}\pi^{-}$ versus $\pi^{+}\pi^{-}\pi^{0}$]{\includegraphics[width=0.4\textwidth,height=0.22\textheight]{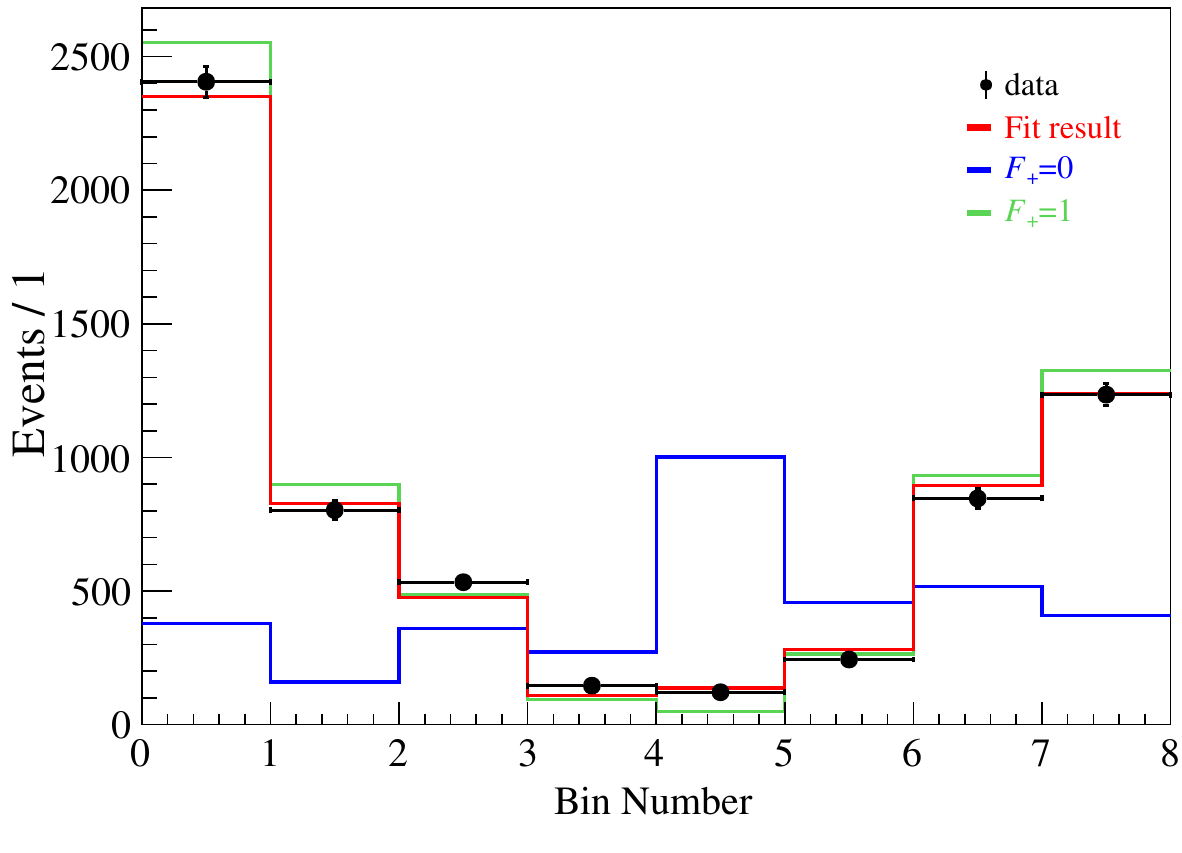}}
		\subfigure[~$K_{S}^{0}\pi^{+}\pi^{-}$ versus $K^{+}K^{-}\pi^{0}$]{\includegraphics[width=0.4\textwidth,height=0.22\textheight]{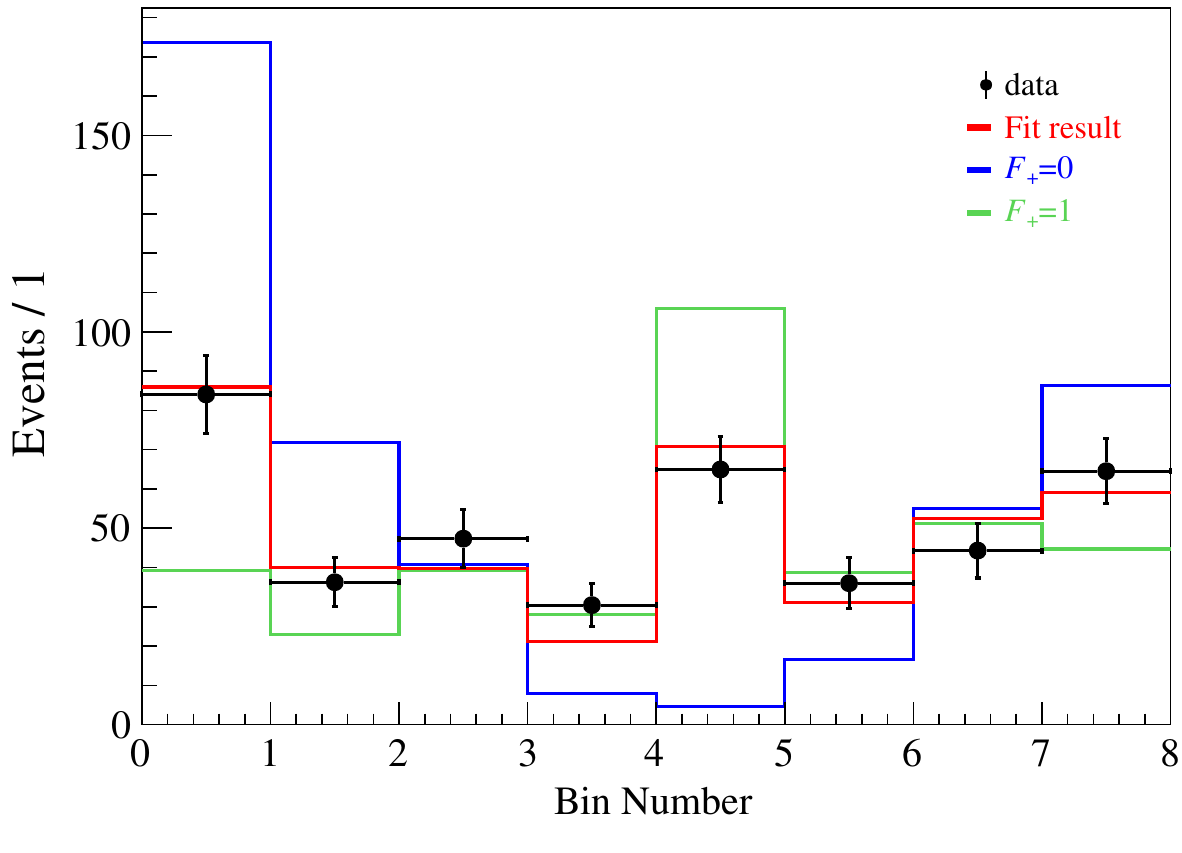}}
		\subfigure[~$K_{L}^{0}\pi^{+}\pi^{-}$ versus $K^{+}K^{-}\pi^{0}$]{\includegraphics[width=0.4\textwidth,height=0.22\textheight]{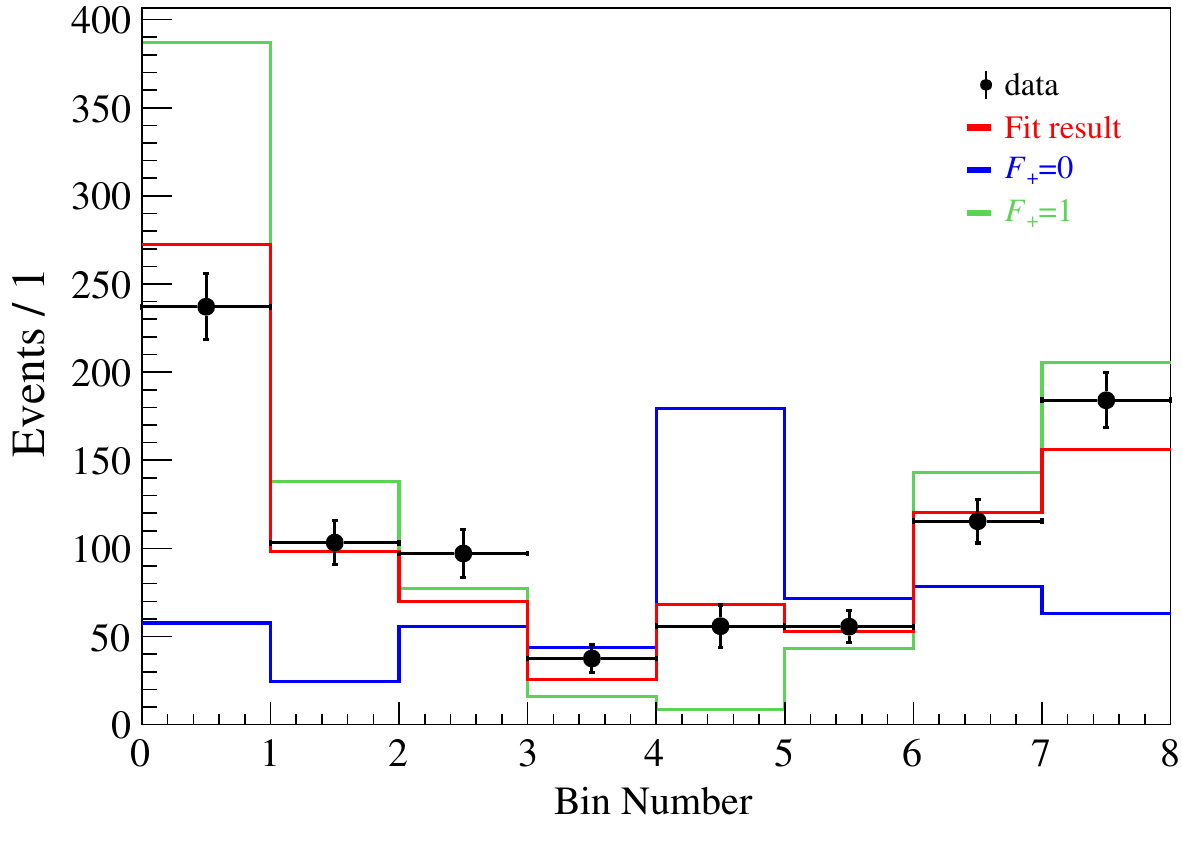}}
		
	\end{center}
	\caption{
		The fit results from the tag modes  $D\to K_{S,L}^{0}\pi^{+}\pi^{-}$ for $D\to \pi^{+}\pi^{-}\pi^{0}$ (a, b) and $D\to K^{+}K^{-}\pi^{0}$ (c, d), respectively. The black points with error bars show the measured values in each bin. The red lines show the predicted values from the fit. The blue and green lines represent the predicted values under the hypotheses of $F_{+}=0$ and $F_{+}=1$, respectively.
	} \label{fig:Kslpipi}
\end{figure*}

Accounting for migration effects, the expected DT yields in Eqs.~\ref{eq:NDT_MixCP_Kspipi} and~\ref{eq:NDT_MixCP_Klpipi} become
\begin{equation} 
	\label{eq:NDT_MixCP_Kspipi_Mig}
	M_{i}=h\sum_{j=1}^{8}\epsilon_{ij}[K_{j}+K_{-j}-2\sqrt{K_{j}K_{-j}}c_{j}(2F^{g}_{+}-1)],
\end{equation}
\noindent
for the tag mode  $D\to K^{0}_{S}\pi^{+}\pi^{-}$  and 
\begin{equation} 
	\label{eq:NDT_MixCP_Klpipi_Mig}
	M'_{i}=h'\sum_{j=1}^{8}\epsilon'_{ij}[K'_{j}+K'_{-j}+2\sqrt{K'_{j}K'_{-j}}c'_{j}(2F^{g}_{+}-1)]
\end{equation}
\noindent
for the tag mode $D\to K^{0}_{L}\pi^{+}\pi^{-}$ where $h$ ($h'$) is a normalization factor. 
To extract $F_{+}^s$ for the signal decay, a likelihood fit is performed by minimizing  
\begin{equation} 
	\label{eq:LL_Kslpipi}
	\begin{aligned}
		-2\mathrm{ln}~\mathcal{L}=-2\sum_{i=1}^{8}\mathrm{ln}~G(M^{\mathrm{obs}}_{i},\sigma_{M^{\mathrm{obs}}_{i}};M^{\mathrm{exp}}_{i}) \\ -2\sum_{i=1}^{8}\mathrm{ln}~G(M'^{\mathrm{obs}}_{i},\sigma'_{M'^{\mathrm{obs}}_{i}};M'^{\mathrm{exp}}_{i}),
	\end{aligned}
\end{equation}
\noindent
where $G$ is a Gaussian function, $M^{(\prime)\mathrm{obs}}_{i}$ is the observed DT yield with peaking background subtracted, $M^{(\prime)\mathrm{exp}}_{i}$ is the expected DT yield and $\sigma_{M^{(\prime)\mathrm{obs}}_{i}}$ is the uncertainty of the DT yield in the $i^{\rm th}$ phase-space bin. 
Individual and simultaneous fits on the  $D\to K^{0}_{S}\pi^{+}\pi^{-}$ and $D\to K^{0}_{L}\pi^{+}\pi^{-}$ tag modes are carried out, which yield the results that are summarized in Table~\ref{tab:fsresut}.
The measured and expected DT signal yields (based on the simultaneously fit results) in individual bins are shown in Fig.~\ref{fig:Kslpipi}. In addition, the expected DT signal yields in individual phase-space  bins under the hypotheses $F_{+}=0$ and 1 are also shown in this figure.

\begin{table}[!htbp]
	\centering
	\caption{Summary of  $F_{+}^{g}$ for the tag modes ${D\to K^{0}_{S, L}\pi^{+}\pi^{-}}$ from individual and simultaneous fits.}	
	\label{tab:fsresut}
	\footnotesize
	\begin{tabular}[b]{ p{4cm} p{4cm} p{4cm} }
		\hline
		\hline
		\multicolumn{1}{l}{Signal/Tag mode}            & \multicolumn{1}{c}{~~~~~$D\to \pi^+\pi^-\pi^0 $~~~~~} & \multicolumn{1}{c}{~~~~~$D\to K^+K^-\pi^0 $~~~~~}  \\
		\hline
          \multicolumn{1}{l}{$D\to K^{0}_{S}\pi^{+}\pi^{-}$} &   \multicolumn{1}{c}{$0.885\pm 0.019$}  & \multicolumn{1}{c}{$0.663\pm0.048$}    \\ 
          \multicolumn{1}{l}{$D\to K^{0}_{L}\pi^{+}\pi^{-}$} &   \multicolumn{1}{c}{$0.920\pm 0.015$}  & \multicolumn{1}{c}{$0.643\pm0.044$}   \\
          \multicolumn{1}{l}{Combined                     } &   \multicolumn{1}{c}{$0.907\pm 0.012$}  & \multicolumn{1}{c}{$0.652\pm0.033$}   \\
		\hline	
		\hline
	\end{tabular}
\end{table}

\section{SYSTEMATIC UNCERTAINTIES ON THE $\textit{\textbf{CP}}$-EVEN FRACTION}
\label{sec:syst_uncer}

All systematic uncertainties of the $F_{+}$ measurement are discussed below, and summarized in Table~\ref{tab:Syst_pipipi0}.

From inspection of Eqs.~\ref{eq:Fp_PureCP} and \ref{eq:eq4}, it can be seen that the systematic uncertainties in the $CP$-even fraction measurement with the $CP$-eigen fully reconstructed tag modes consists of those associated with the ST and DT yields,  the ratio between the DT and ST detection efficiencies, and the factor of $1\pm y$. Since the value of $y$ is ${\cal O}(10^{-2})$ and known with a precision of 5\%~\cite{HFLAV:y}, it introduces negligible uncertainty into the analysis. 
The ratio between the DT and ST detection efficiencies is the efficiency of detecting the signal ($D\to\pi^+\pi^-\pi^0$ and $D\to K^+ K^-\pi^0$), which is almost the same for all tag modes. Therefore the corresponding detection uncertainties cancel in the $F_{+}^g$ calculation. 
The uncertainties associated with the ST and DT yields include those associated with the fit procedure and peaking-background estimation.
The uncertainties associated with the fit procedure are estimated by floating the end point of the ARGUS function, which is fixed in the baseline fits, and taking the resultant differences in the yields with respect to the baseline values as the uncertainties. 
The uncertainties associated with the peaking backgrounds are studied by varying their branching fractions according to the uncertainties recorded in the PDG~\cite{PDG2023}.

For the $CP$-eigen partially reconstructed tag modes,  $N^{\pm}$ is also determined using Eq.~\ref{eq:Fp_PureCP}. Therefore, the sources of systematic uncertainty are the same as those in the fully reconstructed tag modes, and the corresponding uncertainties can be estimated accordingly, apart from those associated with the ST yields.
The ST yields are calculated with Eq.~\ref{eq:NST_PureCP}, and their uncertainties  are associated with the total number of $D\bar{D}$ pairs  $N_{D\bar{D}}$ of the data sample, the branching fractions of $D\to K^{0}_{L}X$ and the tag detection efficiencies.
The uncertainty of $N_{D\bar{D}}$ is taken from Refs.~\cite{BESIII:NDD1,BESIII:NDD2,BESIII:NDD3} and those associated with the branching fractions of $D\to K^{0}_{L}X$ are taken from the PDG~\cite{PDG2023}.
The corresponding uncertainties are studied as follows.
The uncertainties associated with the  $\pi^{\pm}$ tracking and PID are both assigned as 0.5\%,  based on  studies of control samples of ${D\to K^{-}\pi^{+}}$, ${K^{-}\pi^{+}\pi^{-}\pi^{+}}$ versus ${\bar{D}\to K^{+}\pi^{-}}$, ${K^{+}\pi^{-}\pi^{-}\pi^{+}}$ and ${D^{+}\to K^{-}\pi^{+}\pi^{+}}$ versus ${D^{-}\to K^{+}\pi^{-}\pi^{-}}$. 
The uncertainty associated with the  $\pi^{0}$ reconstruction is assigned to be 2\% from studies of the control sample of ${D\to K^{-}\pi^{+}\pi^{0}}$ decays. 
The uncertainties arising from the  ``no extra $\pi^{0}$ and $\eta$" and ``no extra charged tracks" requirements are assigned to be 3\% and 1\% from the analysis of the control samples of ${D\to K^{-}\pi^{+}\pi^{0}}$ and ${D\to \pi^{+}\pi^{-}\pi^{+}\pi^{-}}$ decays. 

The measurements of $F_{+}^{\pi^+\pi^-\pi^0}$ and $F_{+}^{K^+K^-\pi^0}$ made with the pure $CP$ tag  modes are obtained by combining the inputs from $CP$-odd and $CP$-even eigenstates.
To evaluate the overall uncertainty for a specfic uncertainty source, an alternative measurement is performed that includes  the uncertainties in each individual ST tag mode. Then the corresponding uncertainty is calculated by   $\sigma_{\mathrm{syst}}=\sqrt{\sigma_{\mathrm{all}}^{2}-\sigma_{\mathrm{stat}}^{2}}$, where $\sigma_{\mathrm{all}}$ is the total uncertainty after considering the statistical and systematic uncertainties, and their correlations,   and $\sigma_{\mathrm{stat}}$ is  the statistical uncertainty alone.

The global $CP$-mixed tag modes have uncertainties arising from the knowledge of the ST and DT yields and the ratio between DT and ST detection efficiencies. Additional sources of uncertainty come from the input parameters of the $CP$-even fraction $F_{+}^{f}$ and $\langle N^+ \rangle$. 
The systematic uncertainties arising from the DT and ST yields are estimated with the same approaches as used in the measurements with pure $CP$ tag modes.
The detection uncertainties of the ratio between the DT and ST detection efficiencies cancel according to the $N^{f/g}$ and $\langle N^+\rangle$ calculations.
The uncertainty on  $F_{+}^{\pi^+\pi^-\pi^+\pi^-}$ is taken from Ref.~\cite{BESIII:Fp4pi}, while those of  $F_{+}^{\pi^+\pi^-\pi^0}$ and $F_{+}^{K^+ K^-\pi^0}$ are taken from the current analysis.
For the determination using  the identical tag and signal modes, the uncertainty associated with $F_{+}^{g}$ is not considered. The uncertainty in the $F_+^{g}$ measurement associated with the ST and DT yields is obtained through uncertainty propagation according to Eqs.~\ref{eq:Fp_MixCP} and \ref{eq:eq9}. Those associated with $F_+^f$ and $\langle N^+\rangle$ are determined by generating a large number of simulated pseudoexperiments,  where $F_+^f$ and $\langle N^+\rangle$  are sampled from a Gaussian distribution, and the corresponding standard deviations of $F_+^{g}$ are taken as the uncertainties. Here the uncertainties of $\langle N^+\rangle$ include both the statistical and systematic uncertainties of the ST and DT yields.

In the $F_{+}^{g}$ measurement performed in bins of phase-space of the  mixed-$CP$ tag modes $D\to K^{0}_{S,L}\pi^{+}\pi^{-}$, the uncertainties include those associated with the DT yields, the  input parameters of $c_{i}(c'_{i})$ and $K_i(K'_i)$ and bin migration effects.
The uncertainties of the DT yields associated with the same fit procedure are assumed to be identical among different phase-space bins, and therefore cancel in the $F_{+}^{g}$ extraction.
The uncertainties arising from the peaking-background in the DT yields and from the knowledge of the migration matrices are estimated with a `toy MC' method, where fits are performed to a large number of simulated samples in which the size of the peaking background and the elements of the migration matrices are sampled from  Gaussian functions with means and widths set to the values in data.  The  standard deviations of the distributions of fitted $F_{+}^{g}$ values are taken as the corresponding uncertainties. 
The uncertainties associated with the input parameters of $c_{i}$, $c'_{i}$, $K_{i}$ and $K'_{i}$ are estimated with a revised likelihood function given by 
\begin{equation} 
	\label{eq:LL_Kslpipi_input}
	\begin{aligned}
		-2\mathrm{ln}~\mathcal{L}=-2\sum_{i=1}^{8}\mathrm{ln}G(M^{\mathrm{obs}}_{i},\sigma_{M^{\mathrm{obs}}_{i}};M^{\mathrm{exp}}_{i}) \\ -2\sum_{i=1}^{8}\mathrm{ln}G(M'^{\mathrm{obs}}_{i},\sigma'_{M'^{\mathrm{obs}}_{i}};M'^{\mathrm{exp}}_{i})  \\
		-2\mathrm{ln}G(K_{i},K'_{i},\sigma_{K_{i},K'_{i}}) \\
		-2\mathrm{ln}G(c_{i},c'_{i},\sigma_{c_{i},c'_{i}}),
	\end{aligned}
\end{equation}
\noindent
where $G(K_{i},K'_{i},\sigma_{K_{i},K'_{i}})$ ($G(c_{i},c'_{i},\sigma_{c_{i},c'_{i}})$) is the Gaussian function with the means of $K_{i}$ and $K'_{i}$ ($c_{i}$ and $c'_{i}$) taking into account the correlated uncertainty matrix   $\sigma_{K_{i},K'_{i}}$ ($\sigma_{c_{i},c'_{i}}$).  
An alternative fit with Eq.~\ref{eq:LL_Kslpipi_input} is performed with the parameters reported in Ref.~\cite{BESIII:Kslpipi}, the resultant $\sigma=\sqrt{\sigma_{\mathrm{revised}}^{2}-\sigma_{\mathrm{stat}}^{2}}$ is taken as the uncertainty, where $\sigma_{\mathrm{revised}}$ and $\sigma_{\mathrm{stat}}$ are the resultant uncertainties of the fits with the revised and baseline likelihood functions, respectively.

In addition to the above uncertainties, the uncertainty due to MC modeling must be considered.
In the measurements with $CP$-eigen and global $CP$-mixed tag modes, the MC modeling affects the ratio between the DT and ST efficiencies   $\epsilon_{\mathrm{DT}}/\epsilon_{\mathrm{ST}}$, therefore the effects on the tag-side efficiency cancel, and only the MC modeling of signal side ($D\to \pi^+\pi^-\pi^0$ and $D\to K^+K^-\pi^0$) needs to be taken into account. 
To estimate the corresponding uncertainties, a large number (500) of toy MC samples are generated by sampling the modeling parameters of signal side with a Gaussian function incorporating their means and uncertainties obtained from the amplitude analysis.
The measurements are repeated based on the signal MC samples with different modeling parameters, individually, and the standard deviations of $F_+^g$ are taken as the uncertainties. 
In the measurement with the binned mixed-$CP$ tags ($D\to K^{0}_{S,L}\pi^{+}\pi^{-}$), the MC modeling affects efficiencies in different phase-space bins, therefore only the effects on the tag side are considered (the uncertainties on the signal side actually having been already considered as those associated with $c_{i}$, $c'_{i}$, $K_{i}$ and $K'_{i}$).
To estimate the corresponding uncertainty, a conservative estimation with the assumption of uniform tag efficiencies in different phase-space bins is carried out, in which Eqs.~\ref{eq:NDT_MixCP_Kspipi_Syst} and~\ref{eq:NDT_MixCP_Klpipi_Syst} are revised as
\begin{equation} 
	\label{eq:NDT_MixCP_Kspipi_Syst}
	M_{i}=h\sum_{j=1}^{8}\frac{\epsilon_{ij}}{\epsilon^{\mathrm{ST}}_{j}}[K_{j}+K_{-j}-2\sqrt{K_{j}K_{-j}}c_{j}(2F^{g}_{+}-1)],
\end{equation}
\noindent
\begin{equation} 
	\label{eq:NDT_MixCP_Klpipi_Syst}
	M'_{i}=h'\sum_{j=1}^{8} \frac{\epsilon'_{ij}}{\epsilon'^{\mathrm{ST}}_{j}}[K'_{j}+K'_{-j}+2\sqrt{K'_{j}K'_{-j}}c'_{j}(2F^{g}_{+}-1)].
\end{equation}
\noindent
where $\epsilon_{j}^{\mathrm{ST}}$ ($\epsilon'^{\mathrm{ST}}_{j}$) is the tag detection efficiency for the events produced in the $j^{\rm th}$ phase-space bin. The resultant difference in $F^{g}_{+}$ is taken as the systematic uncertainty associated with the signal MC modeling. 

\begin{table*}[!htbp]
	\centering
	\caption{Summary of the systematic uncertainties for ${D^0\to \pi^{+}\pi^{-}\pi^{0}}$ and ${D^0\to K^{+}K^{-}\pi^{0}}$.}	
	\label{tab:Syst_pipipi0}
	\begin{tabular}[b]{c | p{1.7cm} | p{1.7cm} | p{1.7cm} | p{1.7cm} | p{1.7cm}}
		\hline
		\hline
		\small{Source} & \makecell[c]{\small{CP eigen}} & \makecell[c]{\small{$K^{+}K^{-}\pi^{0}$}} & \makecell[c]{\small{$\pi^{+}\pi^{-}\pi^{0}$}} & \makecell[c]{\small{2($\pi^{+}\pi^{-}$)}} & \makecell[c]{\small{$K^{0}_{S,L}\pi^{+}\pi^{-}$}} \\  \hline
		\hline
              \multicolumn{6}{c}{$D\to \pi^{+}\pi^{-}\pi^0$} \\   \hline
		
		Migration  & \makecell[c]{---} & \makecell[c]{---} & \makecell[c]{---} & \makecell[c]{---} & \makecell[c]{$0.0010$} \\
		
		\hline
		ST/DT  & \makecell[c]{$0.0022$} & \makecell[c]{$0.0133$} & \makecell[c]{$0.0081$} & \makecell[c]{$0.0141$} &  \makecell[c]{0.0016} \\
		\hline
		Inputs & \makecell[c]{---} & \makecell[c]{$0.0354$} & \makecell[c]{$0.0006$} & \makecell[c]{$0.0226$} & \makecell[c]{0.0070}   \\
		\hline
		MC model & \makecell[c]{$0.0004$} & \makecell[c]{$0.0030$} & \makecell[c]{$0.0001$} & \makecell[c]{$0.0016$} & \makecell[c]{$0.0003$} \\
		\hline
		Total & \makecell[c]{$0.0022$} & \makecell[c]{$0.0379$} & \makecell[c]{$0.0081$} & \makecell[c]{$0.0267$} & \makecell[c]{$0.0073$} \\
		\hline
		\hline
               \multicolumn{6}{c}{$D\to K^{+}K^{-}\pi^0$}            \\   \hline
		Migration   & \makecell[c]{---} & \makecell[c]{---} & \makecell[c]{---} & \makecell[c]{---} & \makecell[c]{$0.001$} \\
		\hline
		ST/DT  & \makecell[c]{$0.006$} & \makecell[c]{$0.009$} & \makecell[c]{$0.012$} & \makecell[c]{$0.016$} & \makecell[c]{0.002} \\
		\hline
		Inputs & \makecell[c]{---} & \makecell[c]{$0.023$} & \makecell[c]{$0.016$} & \makecell[c]{$0.028$} & \makecell[c]{0.006} \\
		\hline
		MC model  & \makecell[c]{$0.005$} & \makecell[c]{$0.004$} & \makecell[c]{$0.001$} & \makecell[c]{$0.004$} & \makecell[c]{$0.022$} \\
		\hline
		Total & \makecell[c]{$0.008$} & \makecell[c]{$0.025$} & \makecell[c]{$0.020$} & \makecell[c]{$0.033$} & \makecell[c]{$0.023$} \\
		\hline	
		\hline
	\end{tabular}
\end{table*}

\section{COMBINATION OF RESULTS}
\label{sec:combine}

The  results of $F_{+}^g$ for the different categories of tags are summarized in Table~\ref{tab:Fp_sum}.
Least $\chi^{2}$ fit, taking into account the correlated and uncorrelated uncertainties among the different tags, is performed to obtain the average value and the $\chi^{2}$ is given by
\begin{equation} 
	\label{eq:chi2_fit}
	\chi^{2}=\Delta F_{+}^{T}V^{-1}\Delta F_{+},
\end{equation}
where $\Delta F_{+}$ is the difference between measured $F_{+}$ and expected value for each category of tags. $V$ is the covariance matrix defined as below,
\begin{equation} 
	\label{eq:err_fit}
	V_{ij}=\left\{ 
		\begin{array}{lc} 
			(\sigma_{i}^{\mathrm{syst.}})^{2}+(\sigma_{i}^{\mathrm{stat.}})^{2}, & i=j \\ 
			\sigma_{i}^{\mathrm{corr.}}\rho_{ij}\sigma_{j}^{\mathrm{corr.}}, & i\neq j\\ 
		\end{array}
			\right.,
\end{equation}
where the index $i(j)$ represents the $i(j)$-th category of tags, $\sigma_{i}^{\mathrm{syst.}}$ and $\sigma_{i}^{\mathrm{stat.}}$ are the total systematic and statistical uncertainty for $i$-th category of tags. $\sigma_{i(j)}^{\mathrm{corr.}}$ is the correlation uncertainties of $i(j)$-th category of tags and $\rho_{ij}$ is the correlation coefficients among $i$-th and $j$-th categories of tags. The correlation coefficients of the obtained $F_{+}$ under the different tag modes, as summarized in  Table~\ref{tab:correlation_sum_pipipi0}, mainly arise from the common inputs of $\langle N^+ \rangle$, which are estimated with toy MC studies with Gaussian functions of each parameter. The results are also shown in Table~\ref{tab:Fp_sum} and the $\chi^2$ per degree of freedom is  2.39 for the $F_+^{\pi^+\pi^-\pi^0}$ combination and 0.21 for the  $F_+^{K^+K^-\pi^0}$ combination.  The worse fit quality in the $D^0 \to\pi^{+}\pi^{-}\pi^{0}$ case is driven by a 2.5~$\sigma$ difference in result between the result obtained with the pure $CP$ tags and that obtained with the $D \to K^0_{S,L}\pi^+\pi^-$ tags.

\begin{table}[htbp]
	\centering
	\caption{Results of $F_{+}^{g}$ from different tag modes and the value of the combination, where the first uncertainties are statistical and the second systematic.}	
	\label{tab:Fp_sum}
	
       \begin{tabular}[b]{ c  c  c }
		\hline
		\hline
		\small{Tag mode} & \multicolumn{1}{c}{~~~~~~~~~~$F_{+}^{\pi^{+}\pi^{-}\pi^{0}}$~~~~~~~~~~} & \multicolumn{1}{c}{~~~~~~~~~~$F_{+}^{K^{+}K^{-}\pi^{0}}$~~~~~~} \\
		\hline
		
		Pure $CP$  & \multicolumn{1}{c}{$0.9432\pm0.0040\pm0.0022$} & \multicolumn{1}{c}{$0.623\pm0.020\pm0.008$} \\
		
		$K^{+}K^{-}\pi^{0}$  & \multicolumn{1}{c}{$1.0060\pm0.0675\pm0.0379$} & \multicolumn{1}{c}{$0.649\pm0.125\pm0.025$} \\
		
		$\pi^{+}\pi^{-}\pi^{0}$  & \multicolumn{1}{c}{$0.9472\pm0.0139\pm0.0081$}&   \multicolumn{1}{c}{$0.631\pm0.030\pm0.020$}  \\
		
		$2(\pi^{+}\pi^{-})$  & \multicolumn{1}{c}{$0.9948\pm0.0230\pm0.0267$}& \multicolumn{1}{c}{$0.667\pm0.058\pm0.033$} \\
		
		$K^{0}_{S,L}\pi^{+}\pi^{-}$  & \multicolumn{1}{c}{$0.9065\pm0.0116\pm0.0073$} & \multicolumn{1}{c}{$0.652\pm0.033\pm0.023$}  \\ \hline
		
		Combined & \multicolumn{1}{c}{$0.9406\pm0.0036\pm0.0021$} & \multicolumn{1}{c}{$0.631\pm0.014\pm0.011$} \\		
		\hline	
		\hline
	\end{tabular}
	
\end{table}

\begin{table}[htbp]
	\centering
	\caption{Correlation coefficients of the obtained $F_+^g$ uncertainties under the different tag modes which mainly arise from the common inputs of $\langle N^+ \rangle$ for the signal decays of $D^{0}\to\pi^{+}\pi^{-}\pi^{0}$ and $D^{0}\to K^{+}K^{-}\pi^{0}$.}	
	\label{tab:correlation_sum_pipipi0}
	
	\begin{tabular}[b]{ p{2.0cm} p{2.0cm} p{1.8cm}p{1.8cm}  }
		\hline
		\hline
		\small{Tag mode($i$)} 
		& \small{Tag mode($j$)}                                                            & \multicolumn{1}{c}{$\rho_{ij}^{\pi^{+}\pi^{-}\pi^{0}}$~} & \multicolumn{1}{c}{~$\rho_{ij}^{K^{+}K^{-}\pi^{0}}$} \\
		\hline
		
		\multicolumn{1}{c}{$CP$ tag} & \multicolumn{1}{c}{$K^{+}K^{-}\pi^{0}$}                               & \multicolumn{1}{c}{$0.031$}  & \multicolumn{1}{c}{$0.106$} \\
		
		\multicolumn{1}{c}{$CP$ tag} & \multicolumn{1}{c}{$\pi^{+}\pi^{-}\pi^{0}$}                            & \multicolumn{1}{c}{$0.015$} & \multicolumn{1}{c}{$0.292$} \\
		
		\multicolumn{1}{c}{$CP$ tag} & \multicolumn{1}{c}{$2(\pi^{+}\pi^{-})$}                   & \multicolumn{1}{c}{$0.050$}  & \multicolumn{1}{c}{$0.234$}\\
		
		\multicolumn{1}{c}{$K^{+}K^{-}\pi^{0}$} & \multicolumn{1}{c}{$\pi^{+}\pi^{-}\pi^{0}$}             & \multicolumn{1}{c}{$0.014$}  & \multicolumn{1}{c}{$0.077$} \\
		
		\multicolumn{1}{c}{$K^{+}K^{-}\pi^{0}$} & \multicolumn{1}{c}{$2(\pi^{+}\pi^{-})$}    & \multicolumn{1}{c}{$0.038$} & \multicolumn{1}{c}{$0.057$}  \\
		
		\multicolumn{1}{c}{$\pi^{+}\pi^{-}\pi^{0}$} & \multicolumn{1}{c}{$2(\pi^{+}\pi^{-})$} & \multicolumn{1}{c}{$0.015$} & \multicolumn{1}{c}{$0.163$} \\
		
		\hline	
		\hline
	\end{tabular}
	
\end{table}

\section{SUMMARY}

In summary, the $CP$-even fractions of $D^0 \to\pi^{+}\pi^{-}\pi^{0}$ and $D^0 \to K^{+}K^{-}\pi^{0}$ are measured by using an $e^{+}e^{-}$ collision data sample corresponding to an integrated luminosity of 7.93 $\mathrm{fb}^{-1}$ collected at the center-of-mass energy 3.773 GeV with the BESIII detector. The results are ${F_+^{\pi^{+}\pi^{-}\pi^{0}}=0.9406\pm0.0036\pm0.0021}$ and ${F_+^{K^{+}K^{-}\pi^{0}}=0.631\pm0.014\pm0.011}$, respectively. 
These are consistent with the previous results performed with CLEO-c data~\cite{CLEO2015a,CLEO2015b} within $1.9~\sigma$ and $1.7~\sigma$, and the precision is improved by factors of 3.9 and 2.6  for ${D\to\pi^{+}\pi^{-}\pi^{0}}$ and ${D\to K^{+}K^{-}\pi^{0}}$, respectively. For ${D\to\pi^{+}\pi^{-}\pi^{0}}$, the uncertainty of $F_{+}$ is mainly due to $N^{-}$. 

Comparing the results among different tag modes, the pure $CP$ tags are the most powerful.  The self-tag modes and the binned $ D\to K^0_{S,L}\pi^+\pi^-$ modes also have a high weight in the combination.
In future studies,  the  $CP$-mixed tag modes of ${D\to 2(\pi^+\pi^-)}$,   ${D\to\pi^+\pi^-\pi^0}$ and  ${D\to K^+ K^-\pi^0}$ can be analysed in bins of phase space to extract $F_+^{s}$ in the same manner as for the $D \to K^0_{S,L}\pi^+\pi^-$ tags.  This approach is expected to lead to further improvements in precision, as will the analysis of the larger $D\bar{D}$ sample of 20~${\rm fb}^{-1}$ now available at BESIII. 
Our measurements of the $CP$-even fractions of ${D\to\pi^{+}\pi^{-}\pi^{0}}$ and ${D\to K^{+}K^{-}\pi^{0}}$ measured provide valuable input for the measurements of the CKM angle $\gamma$ and the search for indirect $CP$ violation in charm-mixing~\cite{LHCb2016a} at the LHCb and Belle-II experiments.

\section{ACKNOWLEDGEMENT}

\input{./acknowledgement_2024-04-30.tex}

\section{APPENDIX: EFFICIENCY MATRICES}
\label{app:eff_matrix}

The efficiency matrices of ${D\to \pi^{+}\pi^{-}\pi^{0}}$ versus ${\bar{D}\to K^{0}_{S,L}\pi^{+}\pi^{-}}$ and ${D\to K^{+}K^{-}\pi^{0}}$ versus ${\bar{D}\to K^{0}_{S,L}\pi^{+}\pi^{-}}$ are shown in TABLE~\ref{tab:effmatrix_Kslpipi}.

\begin{table*}[!htbp]
	\centering
	\caption{Efficiency matrices $\epsilon_{ij}$$~(\%)$ for $\pi^{+}\pi^{-}\pi^{0}$ versus $K^{0}_{S,L}\pi^{+}\pi^{-}$ and $K^{+}K^{-}\pi^{0}$ versus $K^{0}_{S,L}\pi^{+}\pi^{-}$. The row $i$ shows the reconstructed bin and the column $j$ gives the produced bin.}	
	\label{tab:effmatrix_Kslpipi}
	
	\begin{tabular}[b]{ c c c c c c c c c }
		\hline
		\hline
		\small{\makebox[2cm][c]{Bin}} & \makebox[1.8cm][c]{1} & \makebox[1.8cm][c]{2} & \makebox[1.8cm][c]{3} & \makebox[1.8cm][c]{4} & \makebox[1.8cm][c]{5} & \makebox[1.8cm][c]{6} & \makebox[1.8cm][c]{7} & \makebox[1.8cm][c]{8}   \\
		\hline
		\multicolumn{9}{c}{$\epsilon_{ij}$ for $K^{0}_{S}\pi^{+}\pi^{-}$ versus $\pi^{+}\pi^{-}\pi^{0}$} \\
		
		1 & 12.383 & 1.167 & 0.067 & 0.029 & 0.040 & 0.032 & 0.060 & 1.171 \\
		
		2 & 0.685 & 13.551 & 0.405 & 0.008 & 0.000 & 0.011 & 0.017 & 0.066 \\
		
		3 & 0.057 & 0.604 & 15.332 & 0.362 & 0.002 & 0.004 & 0.010 & 0.028 \\
		
		4 & 0.014 & 0.013 & 0.255 & 14.879 & 0.098 & 0.005 & 0.014 & 0.011 \\
		
		5 & 0.069 & 0.008 & 0.004 & 0.309 & 14.232 & 0.628 & 0.011 & 0.009 \\
		
		6 & 0.027 & 0.004 & 0.000 & 0.007 & 0.299 & 12.730 & 0.525 & 0.017 \\
		
		7 & 0.066 & 0.035 & 0.009 & 0.004 & 0.013 & 0.835 & 12.357 & 0.997 \\
		
		8 & 0.963 & 0.118 & 0.019 & 0.004 & 0.017 & 0.045 & 0.928 & 12.121 \\
		
		\multicolumn{9}{c}{$\epsilon_{ij}$ for $K^{0}_{L}\pi^{+}\pi^{-}$ versus $\pi^{+}\pi^{-}\pi^{0}$} \\
		
		1 & 20.198 & 1.537 & 0.018 & 0.000 & 0.000 & 0.114 & 0.044 & 2.107 \\
		
		2 & 0.650 & 19.354 & 0.681 & 0.003 & 0.000 & 0.004 & 0.000 & 0.013 \\
		
		3 & 0.006 & 0.480 & 19.839 & 0.892 & 0.041 & 0.016 & 0.000 & 0.000 \\
		
		4 & 0.000 & 0.000 & 0.204 & 19.637 & 0.355 & 0.000 & 0.003 & 0.000 \\
		
		5 & 0.000 & 0.005 & 0.017 & 0.408 & 19.373 & 0.451 & 0.000 & 0.000 \\
		
		6 & 0.012 & 0.002 & 0.000 & 0.000 & 0.945 & 17.801 & 0.824 & 0.016 \\
		
		7 & 0.026 & 0.000 & 0.000 & 0.028 & 0.000 & 2.207 & 18.237 & 1.660 \\
		
		8 & 1.199 & 0.001 & 0.000 & 0.000 & 0.000 & 0.003 & 2.164 & 17.486 \\
		
		\multicolumn{9}{c}{$\epsilon_{ij}$ for $K^{0}_{S}\pi^{+}\pi^{-}$ versus $K^{+}K^{-}\pi^{0}$} \\
		
		1 & 9.552 & 0.746 & 0.057 & 0.015 & 0.034 & 0.028 & 0.062 & 0.834 \\
		
		2 & 0.433 & 9.981 & 0.306 & 0.004 & 0.000 & 0.005 & 0.016 & 0.033 \\
 		
		3 & 0.027 & 0.404 & 10.809 & 0.285 & 0.002 & 0.003 & 0.008 & 0.026 \\
		
		4 & 0.015 & 0.009 & 0.177 & 10.638 & 0.077 & 0.008 & 0.005 & 0.005 \\
		
		5 & 0.029 & 0.000 & 0.006 & 0.231 & 9.869 & 0.370 & 0.007 & 0.006 \\
		
		6 & 0.012 & 0.000 & 0.005 & 0.002 & 0.207 & 8.863 & 0.362 & 0.018 \\
		
		7 & 0.033 & 0.023 & 0.007 & 0.006 & 0.006 & 0.631 & 8.716 & 0.630 \\
		
		8 & 0.563 & 0.062 & 0.010 & 0.000 & 0.013 & 0.046 & 0.700 & 8.728 \\
		
		\multicolumn{9}{c}{$\epsilon_{ij}$ for $K^{0}_{L}\pi^{+}\pi^{-}$ versus $K^{+}K^{-}\pi^{0}$} \\
		
		1 & 13.474 & 1.163 & 0.002 & 0.000 & 0.016 & 0.041 & 0.052 & 1.414 \\
		
		2 & 0.454 & 13.065 & 0.497 & 0.024 & 0.000 & 0.002 & 0.000 & 0.009 \\
		
		3 & 0.016 & 0.378 & 13.768 & 0.483 & 0.003 & 0.000 & 0.000 & 0.000 \\
		
		4 & 0.000 & 0.005 & 0.257 & 14.327 & 0.107 & 0.000 & 0.000 & 0.001 \\
		
		5 & 0.005 & 0.000 & 0.008 & 0.147 & 15.329 & 0.453 & 0.006 & 0.001 \\
		
		6 & 0.002 & 0.002 & 0.001 & 0.000 & 0.387 & 12.972 & 0.569 & 0.013 \\
		
		7 & 0.013 & 0.000 & 0.001 & 0.004 & 0.000 & 1.279 & 12.355 & 1.169 \\
		
		8 & 0.854 & 0.000 & 0.007 & 0.000 & 0.000 & 0.006 & 1.360 & 12.018 \\

		\hline	
		\hline
	\end{tabular}
	
\end{table*}

\bibliographystyle{./apsrev4-1}
\bibliography{bibitem}

\nolinenumbers

\end{document}

%% file: authorlist_2024-04-30.tex
\author{
	\begin{small}
		\begin{center}
			M.~Ablikim$^{1}$, M.~N.~Achasov$^{4,c}$, P.~Adlarson$^{76}$, O.~Afedulidis$^{3}$, X.~C.~Ai$^{81}$, R.~Aliberti$^{35}$, A.~Amoroso$^{75A,75C}$, Q.~An$^{72,58,a}$, Y.~Bai$^{57}$, O.~Bakina$^{36}$
% ...
			\\
			\vspace{0.2cm}
			(BESIII Collaboration)\\
			\vspace{0.2cm} {\it
				$^{1}$ Institute of High Energy Physics, Beijing 100049, People's Republic of China\\
% ...
$^{3}$ Bochum  Ruhr-University, D-44780 Bochum, Germany\\
$^{4}$ Budker Institute of Nuclear Physics SB RAS (BINP), Novosibirsk 630090, Russia\\
% ...
$^{35}$ Johannes Gutenberg University of Mainz, Johann-Joachim-Becher-Weg 45, D-55099 Mainz, Germany\\
$^{36}$ Joint Institute for Nuclear Research, 141980 Dubna, Moscow region, Russia\\
% ...
$^{57}$ Southeast University, Nanjing 211100, People's Republic of China\\
$^{58}$ State Key Laboratory of Particle Detection and Electronics, Beijing 100049, Hefei 230026, People's Republic of China\\
% ...
$^{72}$ University of Science and Technology of China, Hefei 230026, People's Republic of China\\
% ...
$^{75}$ University of Turin and INFN, (A)University of Turin, I-10125, Turin, Italy; (B)University of Eastern Piedmont, I-15121, Alessandria, Italy; (C)INFN, I-10125, Turin, Italy\\
$^{76}$ Uppsala University, Box 516, SE-75120 Uppsala, Sweden\\
% ...
$^{81}$ Zhengzhou University, Zhengzhou 450001, People's Republic of China\\
\vspace{0.2cm}
$^{a}$ Deceased\\
% ...
$^{c}$ Also at the Novosibirsk State University, Novosibirsk, 630090, Russia\\
% ...
		}\end{center}	
		\vspace{0.4cm}
	\end{small}
}


%% file: acknowledgement_2024-04-30.tex

The BESIII Collaboration thanks the staff of BEPCII, the IHEP computing center and the supercomputing center of the University of Science and Technology of China (USTC) for their strong support. This work is supported in part by National Key R\&D Program of China under Contracts Nos. 2020YFA0406400, 2020YFA0406300, 2023YFA1606000, 2023YFA1609400; National Natural Science Foundation of China (NSFC) under Contracts Nos. 11635010, 11625523, 11735014, 11935015, 11935016, 11935018, 11961141012, 12025502, 12035009, 12035013, 12061131003, 12122509, 12105276, 12192260, 12192261, 12192262, 12192263, 12192264, 12192265, 12221005, 12225509, 12235017, 12361141819; the Chinese Academy of Sciences (CAS) Large-Scale Scientific Facility Program; the CAS Center for Excellence in Particle Physics (CCEPP); Joint Large-Scale Scientific Facility Funds of the NSFC and CAS under Contract No. U1832207, U2032111, U1732263, U1832103; 100 Talents Program of CAS; CAS Youth Team Program under Contract No. YSBR-101; The Institute of Nuclear and Particle Physics (INPAC) and Shanghai Key Laboratory for Particle Physics and Cosmology; German Research Foundation DFG under Contracts Nos. 455635585, FOR5327, GRK 2149; Istituto Nazionale di Fisica Nucleare, Italy; Ministry of Development of Turkey under Contract No. DPT2006K-120470; National Research Foundation of Korea under Contract No. NRF-2022R1A2C1092335; National Science and Technology fund of Mongolia; National Science Research and Innovation Fund (NSRF) via the Program Management Unit for Human Resources \& Institutional Development, Research and Innovation of Thailand under Contracts Nos. B16F640076, B50G670107; Polish National Science Centre under Contract No. 2019/35/O/ST2/02907; The Swedish Research Council; U. S. Department of Energy under Contract No. DE-FG02-05ER41374.

